\documentclass{aa}
\usepackage{graphicx}
\usepackage{txfonts}
\usepackage{natbib}

\bibpunct{(}{)}{;}{a}{}{,} 
\usepackage[colorlinks,linkcolor=blue,anchorcolor=blue,citecolor=blue]{hyperref}

\begin{document}

\title{Unveiling the nonthermal environments of NGC\,3411 and NGC\,5903 through new MeerKAT observations}

\author{T. Pasini\inst{\ref{inst1}} 
\and F. De Gasperin\inst{\ref{inst1}}
\and E. O'Sullivan\inst{\ref{inst2}} 
\and M. Brüggen\inst{\ref{inst3}}
\and M. Gaspari\inst{\ref{inst4}}
\and K. Kolokythas\inst{\ref{inst5}, \ref{inst6}, \ref{inst1}}
\and E. Bulbul\inst{\ref{inst7}}
\and Y. E. Bahar\inst{\ref{inst8}}
}

\authorrunning{T. Pasini et al.}

\institute{INAF - Istituto di Radioastronomia, via P. Gobetti 101, 40129, Bologna, Italy\label{inst1}; \email{thomas.pasini@inaf.it}
\and Center for Astrophysics | Harvard \& Smithsonian, 60 Garden Street, Cambridge, MA 02138, USA\label{inst2};
\and Universität Hamburg, Hamburger Sternwarte, Gojenbergsweg 112, D-21029, Hamburg, Germany\label{inst3};
\and Department of Physics, Informatics and Mathematics, University of Modena and Reggio Emilia, I-41125 Modena, Italy\label{inst4};
\and Centre for Radio Astronomy Techniques and Technologies (RATT), Department of Physics and Electronics, Rhodes University, Makhanda 6140, South Africa\label{inst5};
\and South African Radio Astronomy Observatory, Black River Park North, 2 Fir St, Cape Town, 7925, South Africa\label{inst6};
\and Max Planck Institute for Extraterrestrial Physics, Giessenbachstrasse 1, 85748 Garching, Germany\label{inst7};
\and INAF, Osservatorio di Astrofisica e Scienza dello Spazio, Via Piero Gobetti 93/3, I-40129 Bologna, Italy\label{inst8};
}

\abstract
{Galaxy groups are key laboratories for investigating the interplay between active galactic nucleus (AGN) outbursts and the dynamical processes of the intra-group medium.}
{We aim to characterize the morphology, spectral properties, and magnetic field structure of the diffuse nonthermal radio emission in two peculiar galaxy groups, NGC\,3411 and NGC\,5903, selected from an originally larger sample, in order to constrain its origin and its connection with the dynamics of the intra-group medium.}
{We present new MeerKAT UHF- and L-band observations of the two groups, complemented by archival \textit{Chandra} and \textit{XMM-Newton} data already analyzed in previous works.}
{In NGC\,3411, we detect extended radio emission surrounding the central AGN over scales of $\sim70$ kpc, and resolve previously undetected arc-shaped substructures on scales of $\sim20$--30 kpc. This extended component exhibits a relatively uniform steep spectrum ($\alpha \simeq -1.7$). We discuss possible scenarios for its origin, including mini-halos, aged AGN plasma, weak shocks, and the possibility of a radio galaxy observed along the line of sight. In NGC\,5903, MeerKAT reveals a complex filamentary morphology, including a previously unknown ring-like structure surrounding the central AGN. Spectral-index mapping shows localized flattening ($\alpha \simeq -0.9$) along the north arc, while polarization data reveal ordered magnetic fields and a patchy rotation measure distribution across the source. Together with the disturbed X-ray morphology, these results support a scenario in which fossil AGN plasma has been redistributed and locally reenergized by group-scale dynamical processes such as turbulence, shear, and weak compression.}
{Our results demonstrate that galaxy groups can host a rich variety of nonthermal phenomena, whose properties can only be fully revealed through the combination of sensitive, high-resolution, and multifrequency radio observations.}


\maketitle
\nolinenumbers

\authorrunning{T. Pasini, F. De Gasperin, E. O'Sullivan et al.}


\section{Introduction}

Galaxy groups represent the most common environment for galaxies in the local Universe, containing more than half of the total galaxy population \citep[e.g.,][]{Eke_2004, Tully_2015}. As an intermediate step in the hierarchical formation of structures that bridge the gap between individual galaxies and massive clusters, they play a crucial role in driving galaxy evolution. The majority of baryons in these systems reside in the hot, X-ray-emitting intra-group medium (IGrM), the thermodynamic state of which is determined by the interplay between gravitational processes, radiative cooling, and nongravitational heating \citep[see][for a review]{Sun_2012, Lovisari_2021, Bahar_2024}.

Of the nongravitational processes, feedback from active galactic nuclei (AGNs) is widely accepted as the primary mechanism preventing catastrophic cooling of the hot gas in the centers of groups and clusters \citep{McNamara_2007, Fabian_2012}. In the standard "radio mode" feedback scenario, the central supermassive black hole launches relativistic jets that inflate cavities in the surrounding medium. The energy deposited through these bubbles balances radiative losses, maintaining the global thermal equilibrium. While this cycle is well studied in massive galaxy clusters, AGN feedback is expected to be even more impactful in the shallower gravitational potentials of galaxy groups \citep{Giodini_2010, Oppenheimer_2010}, where the energy output of the AGN can be comparable to the binding energy of the gas. Evidence of this impact is provided by the reduced gas fractions observed in galaxy groups \citep[e.g.,][]{Eckert_2021, Siegel_2026}.

A key, yet often elusive, component of the feedback cycle is the generation of shocks, which can be driven by powerful AGN outbursts propagating through the intra-cluster medium (ICM), as observed, for example, in systems like M87 \citep{Forman_2005}, Hydra A \citep{Nulsen_2005}, and MS 0735.6+7421 \citep{McNamara_2005}. These shocks are efficient at converting kinetic energy into thermal energy, thereby reheating the cooling gas \citep{Fabian_2006}. However, detecting such features in galaxy groups has historically been challenging due to the lower surface brightness of the IGrM and the typically smaller physical scales involved compared to those in massive clusters. Consequently, the precise contribution of shock heating to the thermodynamic budget of galaxy groups remains an open question \citep{Osmond_2004, Panagoulia_2014}.

The nonthermal component of the IGrM, as revealed through radio emission, offers a unique window into the history of galaxy groups \citep[e.g.,][]{Pasini_2021}. Relativistic plasma displaced by subsonic motions (e.g., turbulence, uplift, and shear) and by magnetic draping can be gently compressed, stretched, and reenergized, producing filamentary morphologies and regions of locally flatter radio spectra without the need for strong shocks. For this reason, deep radio observations provide a complementary probe of the AGN heating cycle, mixing and transport processes, and the long-term duty cycle of central engines \citep[e.g.,][]{Brienza_2021, Brienza_2022}. Models that include multiphase condensation and chaotic cold accretion predict intermittent, multidirectional outbursts regulated by the competition between cooling and turbulent eddy-turnover times \citep[e.g.,][]{Gaspari_2013, Gaspari_2018}. In this picture, radio emission is not only a tracer of current jet activity or strong shocks but also a long-lived record of fossil plasma. This plasma can be redistributed, compressed, and magnetically draped by the evolving IGrM, which produces sharp filaments and locally flatter spectra without requiring strong shocks through the thermal gas \citep{Gaspari2012, Yang_2019, Wittor_2020}.
Recent improvements in low-surface-brightness sensitivity, provided by instruments such as MeerKAT and the LOw Frequency ARray \citep{vanHaarlem_2013}, have unveiled new classes of diffuse radio sources, including odd radio circles (ORCs; \citealt{Norris_2021}). While their nature remains uncertain, proposed scenarios range from spherical shock waves generated by powerful events to old AGN lobes reenergized by group-scale disturbances \citep[e.g.,][]{Norris_2021, Shabala_2024, Kumari_2024}. Although only a handful of ORCs are currently known, several new candidates have been identified in recent surveys \citep{Norris_2021, Gupta_2022, Koribalski_2024, Bulbul_2024, Gupta_2025}, including those carried out by LOFAR \citep{DeGasperin_2026}.

In this work, we present new MeerKAT observations of two galaxy groups, NGC\,3411 and NGC\,5903. These systems were selected from a larger sample of groups originally used to investigate the coupling and interactions between AGN outbursts and the thermal phase of the IGrM. According to previous observations, both targets exhibit peculiar, multiphase environments and complex radio morphologies that challenge standard feedback models. NGC\,3411 appears relaxed in X-rays but hosts a puzzling temperature structure with a "cool shell" (\citealt{OSullivan_2007}, hereafter OS07). Conversely, NGC\,5903 displays a complex interaction between neutral hydrogen (HI) filaments, hot gas, and steep-spectrum radio emission (\citealt{OSullivan_2018}, hereafter OS18), suggesting a history of interactions and mixing. By combining our new MeerKAT data with archival X-ray and optical observations, performing a spectral analysis, and examining polarization, we aim to disentangle the origin of the nonthermal emission observed in these two systems.

This paper is organized as follows: in Sect. \ref{sec:calibration} we describe the data reduction and calibration, in Sect. \ref{sec:results} we present the images and results for the two targets, and in Sect. \ref{sec:discussion} we discuss the physical implications of our findings. We assume a $\Lambda$ cold dark matter cosmology with $H_0 = 70$ km s$^{-1}$ Mpc$^{-1}$, $\Omega_m = 0.3,$ and $\Omega_\Lambda = 0.7$. At the redshift of NGC\,3411 ($z \sim 0.015274$), 1$''$ = 0.3 kpc, while for NGC\,5903 ($z \sim 0.008556$), 1$''$ = 0.18 kpc. Throughout this work, we adopt the usual spectral index definition: $S_\nu \propto \nu^{\alpha}$, where $S_\nu$ is the flux density at frequency $\nu$ and $\alpha$ is the spectral index.

\section{Data calibration}
\label{sec:calibration}

\begin{figure*}[t!]
\centering
  \includegraphics[scale=0.44]{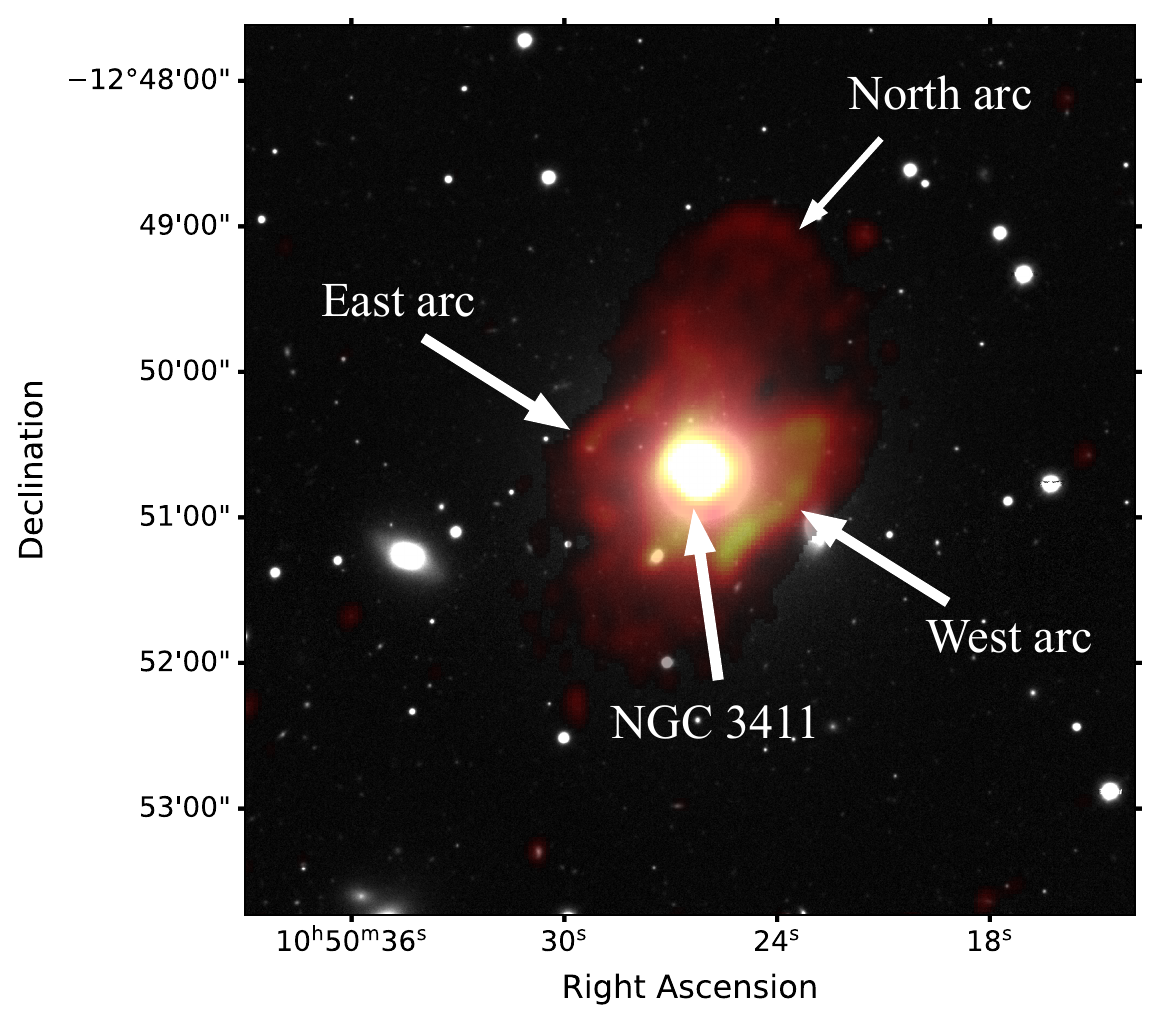}
  \includegraphics[scale=0.452]{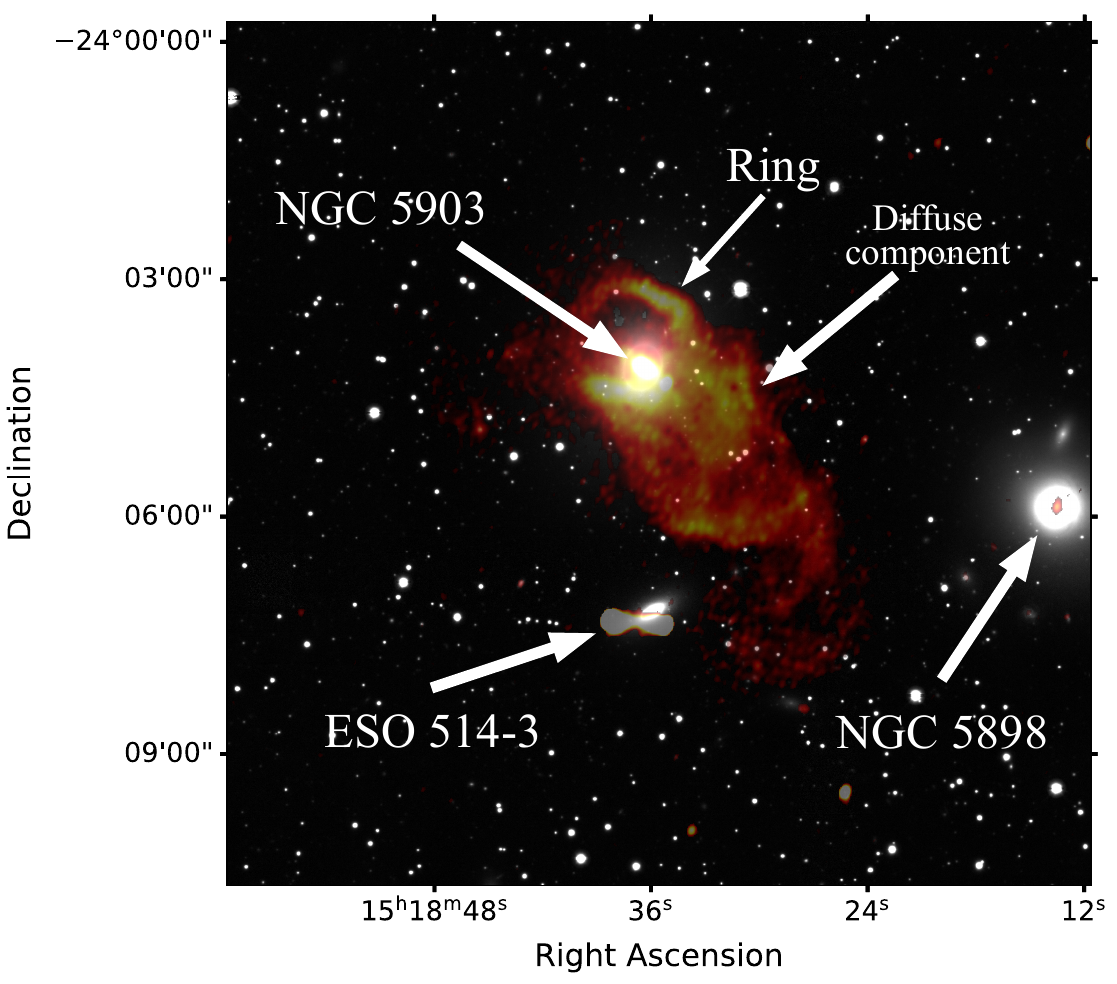}
  \caption{\textit{Left}: Optical image from the DESI Legacy survey of NGC\,3411, with overlaid radio emission (red is fainter, yellow is brighter) as detected with our new UHF-band observation from MeerKAT. \textit{Right}: Optical image from the DESI Legacy survey of NGC\,5903, with overlaid radio emission as detected with our new L-band observation from MeerKAT.}
  \label{fig:overlays}
\end{figure*}

The MeerKAT proprietary data used in this work were obtained in Cycle 5 as part of a dedicated proposal (Principal Investigator T. Pasini, project code SCI-20241101-TP-01), which observed a total of 16 galaxy groups. The sample was initially selected in the X-ray band using the first Data Release of the eROSITA data (eRASS1). We started from the catalog first built and described in \citet{Bahar_2024} and \citet{Bulbul_2024b}, which includes 1178 galaxy groups with $5 \times 10^{12} M_\odot < M_{500} < 10^{14} M_\odot$. The catalog has been accurately cleaned to avoid contamination by, for example, spurious AGNs and is, to date, the largest ICM-selected clean catalog of galaxy groups with available state-of-the-art X-ray data. We then applied the following thresholds: 

\begin{itemize}

\item $10^{13}M_\odot < M_{500} < 10^{14} M_\odot$ and $L_{500} > 10^{42}$ erg s$^{-1}$ (where $L_{500}$ is the X-ray luminosity within $R_{500}$). These were chosen to avoid the lowest-mass systems that could be only barely detected in the X-ray band, which is key to properly understanding the interplay between thermal and nonthermal emission in these systems.

\item $z < 0.03$. This is purely dictated by the required resolution in the radio band: for galaxy groups at $z \sim 0.03$, we are still able to resolve a physical size of $\sim$2 kpc in the L band. This is essential to resolve radio galaxies and to investigate the presence of bubbles, ripples, and filaments that usually indicate a strong interplay with the IGrM \citep[see, e.g.,][]{Timmerman_2022, Brienza_2021}.

\end{itemize}

Through this selection we obtained a total of 16 galaxy groups, which were observed in the L band and UHF band with an on-target exposure of 0.2 hours per target. At first, the data calibration was carried out following the workflow described in \citet{Balboni_2026} and briefly summarized here. We applied the calibration solutions provided by the South African Radio Astronomy Observatory (SARAO) Science Data Processor (SDP)\footnote{\url{https://archive.sarao.ac.za/}}, which correct for delay, bandpass, and gain. We then performed further self-calibration cycles using {\ttfamily facetselfcal\footnote{\url{https://github.com/rvweeren/lofar_facet_selfcal}}} \citep{vanWeeren_2021} to reduce the noise and improve the image quality around the target of interest. Imaging was carried out with {\ttfamily WSClean\footnote{\url{https://gitlab.com/aroffringa/wsclean}}} \citep{Offringa_2014} by applying suitable weighting and tapering of the visibilities in order to obtain images at different resolutions. However, by inspecting the resulting maps, we found a significant flux scale misalignment when comparing our L-band images with the NRAO VLA Sky Survey (NVSS; \citealt{Condon_1998}), and our UHF-band images with the Rapid ASKAP Continuum Survey (RACS; \citealt{McConnell_2020}), even after proper primary beam correction. This misalignment, which resulted in a flux scale consistently higher with respect to aforementioned surveys, was not uniform across the field of view, showed no clear increasing or decreasing gradient with respect to the phase center, and affected both observing bands.
While for a few sources we estimated an increase of $\sim$10 to $\sim$15\%, for others it was as high as 30-50\%, which effectively made the data unusable for science.

We therefore decided to download the raw datasets, without applying any a priori calibration from SARAO, and apply an ad hoc calibration strategy using {\ttfamily MEPI\footnote{\url{https://github.com/revoltek/mepi}}}, a pipeline that was recently developed to carry out continuum and polarization calibration of MeerKAT datasets. We summarize here the main steps. First, calibrator fields (bandpass, phase, and polarization calibrators) are split from the raw measurement set, and a feed-swap correction is applied to account for the known X/Y antenna orientation error. Standard flagging is then performed, and bandpass solutions, antenna-based delays, complex gain solutions, polarization leakage, cross-hand delays, and polarization angle solutions are derived. Calibration tables are applied to the target field, which is further flagged (through {\ttfamily AOFlagger\footnote{\url{https://gitlab.com/aroffringa/aoflagger}}}; \citealt{Offringa_2010}) before being split into a frequency- and time-averaged measurement set. Self-calibration is then performed iteratively. Once the continuum calibration is complete, Stokes Q and U image cubes are produced through {\ttfamily WSClean} and passed to the {\ttfamily RM-Tools\footnote{\url{https://github.com/CIRADA-Tools/RM-Tools}}} \citep{VanEck_2026} package for rotation measure (RM) synthesis, RM-CLEAN \citep{Heald_2009}, and peak-fitting, ultimately yielding maps of the Faraday depth, intrinsic polarization angle, and polarized intensity. Using this method, output continuum images were found to have consistent flux scale with NVSS, for the L band, and RACS, for the UHF band. While it is not trivial to find the origin of the flux scale misalignment in the SARAO calibration, a possibility we propose is saturation from radio frequency interference, which might not have been sufficiently flagged originally, but is dealt with by the more aggressive flagging strategy of {\ttfamily MEPI}. Given these issues, even after fixing the flux scale, for all measurements performed on the resulting images we still assume a conservative flux density uncertainty of $\sim 10\%$ for all bands.

Through visual inspection of our targets, we have identified two galaxy groups whose nonthermal emission exhibits very peculiar properties and morphologies. These are J105026.1-12 and J151828.4-24, which can be identified through their brightest group galaxy (BGG) as NGC\,3411 and NGC\,5903, respectively. In Fig. \ref{fig:overlays} we show for both the radio emission detected through our new MeerKAT observations, overlaid on optical images from the Dark Energy Spectroscopic Instrument (DESI) Legacy Survey\footnote{\url{https://www.legacysurvey.org}} \citep{Dey_2019}. A thorough description of the two systems is given in the next sections. We also carried out polarization calibration for both, using a suitable calibrator (3C286) during the original observations. We followed the workflow of {\ttfamily MEPI}, as described in detail above, focusing on the L-band observations to limit the effects of Faraday depolarization. However, polarized emission was detected only in NGC\,5903, while images for NGC\,3411 remained empty. For this reason, throughout this work, we discuss polarization only for NGC\,5903.

\begin{figure*}[t!]
\centering
  \includegraphics[scale=0.40]{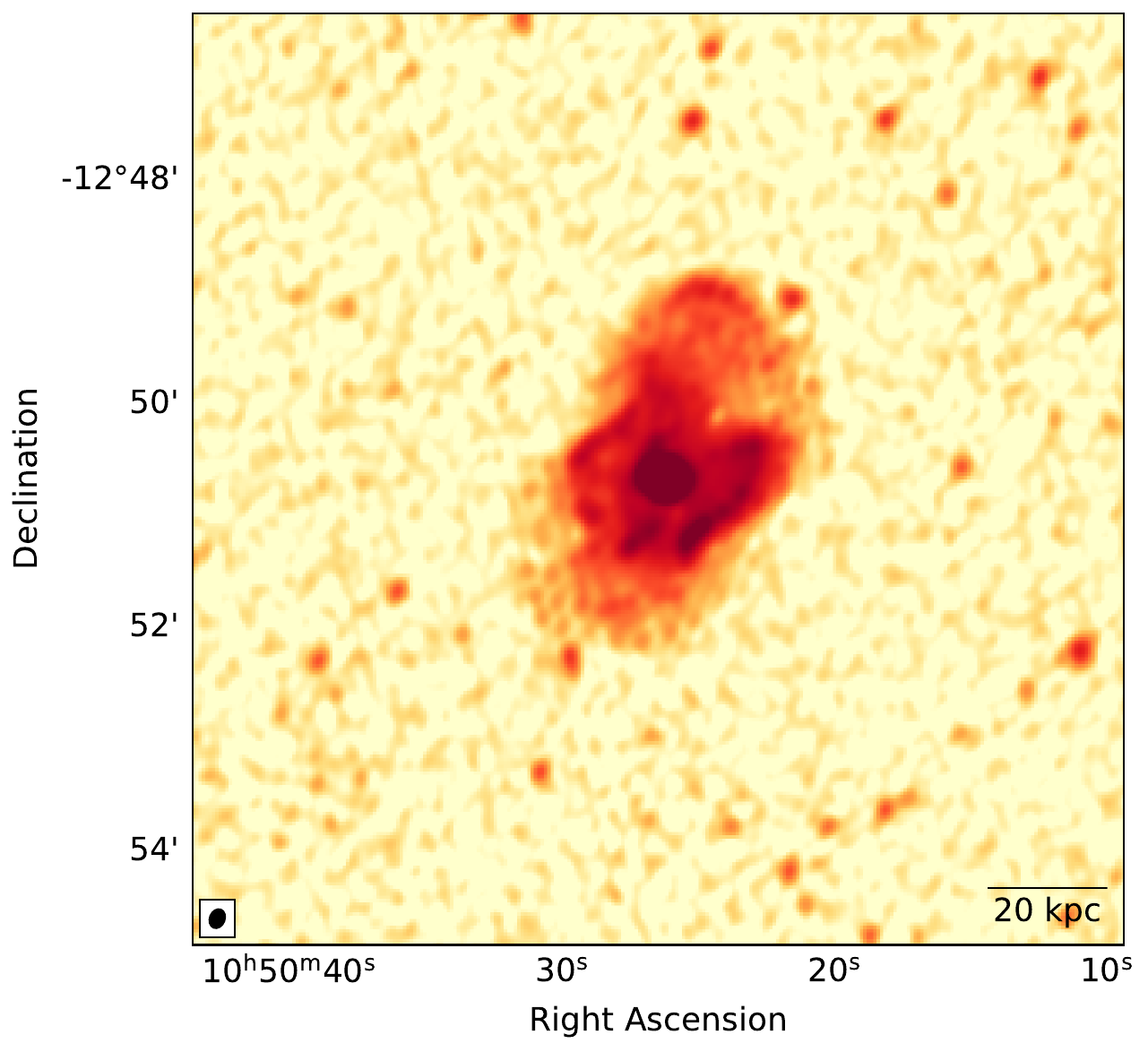}
  \includegraphics[scale=0.40]{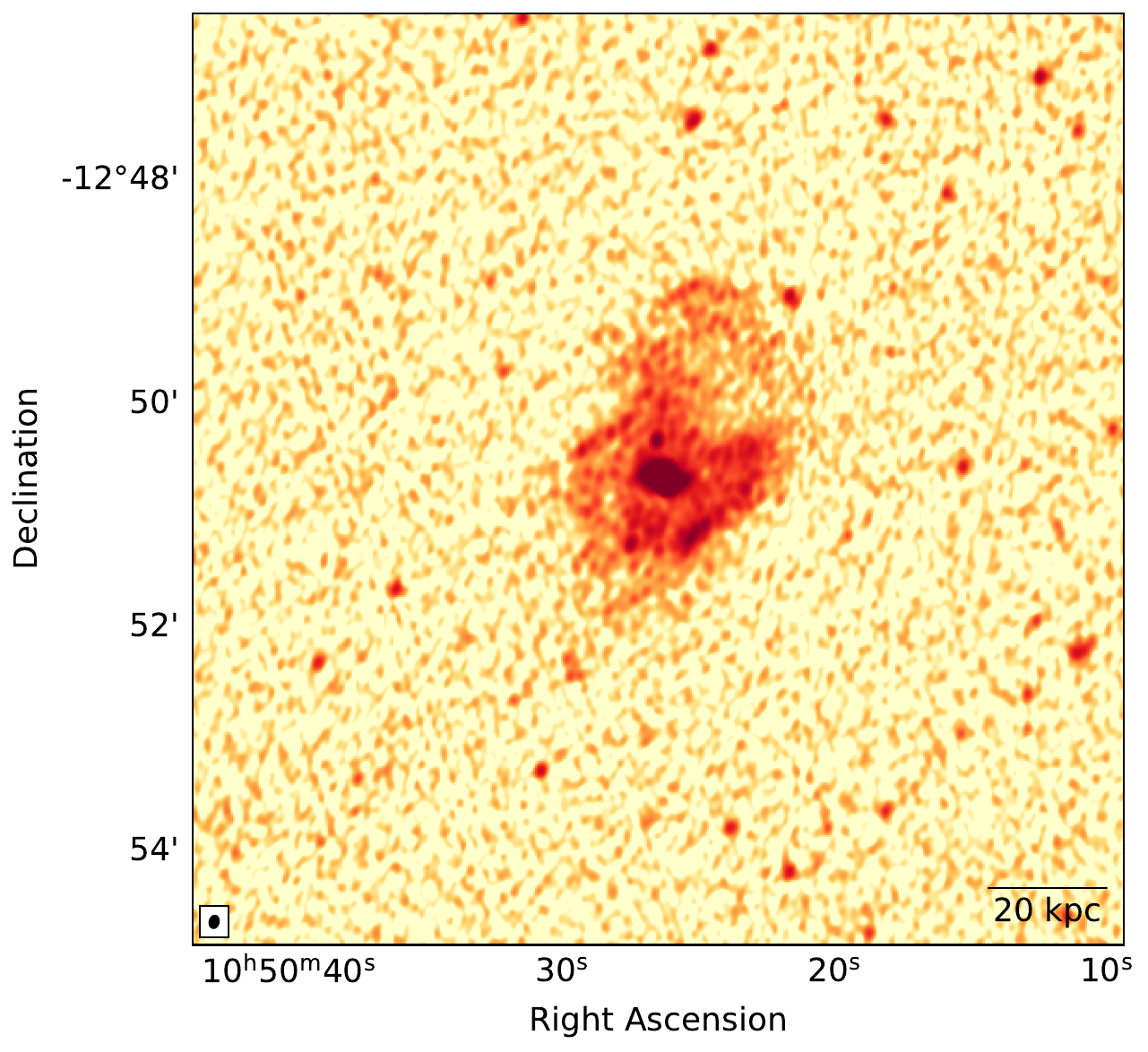}
  \caption{\textit{Left}: UHF-band MeerKAT image of NGC\,3411 produced with {\ttfamily Briggs -0.3}. The rms noise is $\sim$19 $\mu$Jy beam$^{-1}$, and the beam (shown in the bottom left) is $10'' \times 8''$. \textit{Right}: L-band MeerKAT image of NGC\,3411 produced with {\ttfamily Briggs -0.3}. The rms noise is $\sim$18 $\mu$Jy beam$^{-1}$, and the beam (shown in the bottom left) is $7'' \times 5''$.}
  \label{fig:NGC3411}
\end{figure*}

\section{Targets and results}
\label{sec:results}

In this section we describe the two targets of this study and present the results from our MeerKAT observations. We stress that both sources were observed in the UHF and L bands with a total exposure of $\sim$12 minutes per system per band. The central frequency of each dataset depends upon the flags applied during calibration. In our case, for the UHF and L bands, we find 814 and 1280 MHz for NGC\,3411, and 816 and 1284 MHz for NGC\,5903, respectively.

\subsection{NGC\,3411}

\subsubsection{Overview}

NGC\,3411 is the dominant elliptical galaxy of J105026.1-12 (cross-ID USGC S152, SS2b 153, NGC 3402), located at redshift $z \sim$0.015 and with estimated total mass (from velocity dispersion) $\sim 2.7 \times 10^{13}$ M$_\odot$ (see OS07). The system appears relaxed and nearly spherical from X-ray surface brightness maps, making NGC\,3411 a good example of a relatively undisturbed galaxy group environment.

Nevertheless, what makes NGC\,3411 notable is its highly unusual IGrM temperature distribution. Spatially resolved \textit{XMM-Newton} and \textit{Chandra} spectroscopy (see OS07) shows a relatively hot inner region (peak $kT\sim$0.84 keV at $\sim$50 kpc from the center) surrounded by a cooler shell of gas at $\sim$0.7 keV with radius $\sim$20–40 kpc, with the shell being $\sim$0.15 keV cooler than the neighboring gas. Spectral mapping reveals the cool component as an irregular ring. Mass, entropy, and cooling-time profiles are broadly typical for a group of this temperature, but the cooling time falls below $\sim$1 Gyr at $\sim$30 kpc and is $\sim 4 \times 10^8$ yr at $\sim$10 kpc, implying that a sustained heat source is required to prevent runaway cooling.

Radio imaging with the Very Large Array (VLA) at 8, 5, and 1.4 GHz only detects a weak nuclear radio source and a $\sim 15''$ ($\sim$4.5~kpc) blobby diffuse component, without any hint of a more defined structure or morphology. There is no strong optical or X-ray signature of an active AGN, suggesting that the central black hole might currently be quiescent, although it is worth noting that this kind of emission is anyway not expected from elliptical galaxies hosting advection-dominated accretion flow (\citealt{Narayan_1995}) such as the BGG of this group. Using Giant Metrewave Radio Telescope (GMRT) observations, \citet{Giacintucci_2011} detected a much larger ($\sim80$ kpc) diffuse structure at 235 and 610 MHz and highlighted similarities with mini-halos sometimes detected in relaxed cool-core clusters. However, they detect no significant substructure within this diffuse halo.

\subsubsection{The MeerKAT view}

In the left panel of Fig. \ref{fig:NGC3411} we show the UHF-band image of NGC\,3411 obtained from our new observation. This new map reveals a plethora of additional structures and features compared to the older VLA and GMRT images. We detect the central, compact component, which yields a flux density of $S_{814 \rm MHz} = 26 \pm 3$ mJy that translates into a $k$-corrected luminosity of $L_{814 \rm MHz} = (1.3 \pm 0.1) \times 10^{22}$ W Hz$^{-1}$ assuming spectral index $\alpha \sim -0.83$, as estimated in the next section. For the diffuse component, we measure instead a total flux density of $S_{812 \rm MHz} = 61 \pm 6$ mJy within 3$\sigma$ contours, which corresponds to $L_{814 \rm MHz} = (3.1 \pm 0.3) \times 10^{22}$ W Hz$^{-1}$ at the group redshift, excluding the central AGN. Thanks to the high angular resolution of MeerKAT, arc-shaped features are clearly detected east and west of the AGN, embedded within a more extended, low–surface–brightness nonthermal halo. The east arc extends (in projection) for $\sim$30 kpc, while we measure a projected extent of $\sim$25 kpc for the west arc. An additional, $\sim$20 kpc long and fainter arc is detected north of the AGN. The width of all these structures is $\sim$5 kpc\footnote{Fully resolved through our MeerKAT observations.}. None of these features has a clear counterpart in optical (e.g., DESI) or IR catalogs and images (e.g., 2MASS; \citealt{Skrutskie_2006}). The diffuse component shows a complex morphology, with filaments and an asymmetric brightness distribution, suggesting the presence of multiple substructures within the radio-emitting plasma. The overall projected extent of the source is $\sim$50 kpc on the major axis (north-south) and $\sim$30 kpc on the east-west axis at this frequency.

In the right panel of Fig. \ref{fig:NGC3411} we show instead the L-band image of the galaxy group. While the diffuse component is less prominent at this frequency, suggesting a steep spectrum, it is still clearly detected and exhibits very similar features to those previously described for the UHF-band image. The north arc is also evident, while the emission south of the AGN, aside from the brightest region, is faint and gets mixed up with the noise. The flux density of the AGN at this frequency is $S_{1280 \rm MHz} = 18 \pm 2$ mJy, which translates into a luminosity 
$L_{1280 \rm MHz} = (9.1 \pm 0.9) \times 10^{21}$ W Hz$^{-1}$. For the diffuse component we measure $S_{1280 \rm MHz} = 27.4 \pm 2.7$ mJy, which corresponds to $L_{1280 \rm MHz} = (1.4 \pm 0.1) \times 10^{22}$ W Hz$^{-1}$ at the group redshift, excluding the central AGN.

\subsubsection{Spectral index map}

Assuming the luminosity values reported in the previous section, we can estimate for the central AGN an integrated spectral index $\alpha_{814 \rm MHz}^{1280 \rm MHz} = -0.83 \pm 0.31$, and $\alpha_{814 \rm MHz}^{1280 \rm MHz} = -1.75 \pm 0.31$ for the diffuse halo. However, the peculiar morphology and substructures revealed by MeerKAT suggest that an accurate understanding of the origin of this source would require disentangling the various morphological components.
We therefore produced a spectral index map by combining the MeerKAT UHF- and L-band observations. First, maps at the two frequencies were produced by applying the same visibility cut 125$\lambda$ - 21k$\lambda$, using {\ttfamily Briggs -0.3} weighting, and convolving them at a common beam $10'' \times 8''$. After re-gridding, the spectral index and related error maps were then generated by calculating $\alpha$ and $\Delta\alpha$ in each pixel following
\begin{equation}
\label{eq:spindex}
\alpha_{\nu_1}^{\nu_2} \pm \Delta \alpha_{\nu_1}^{\nu_2} = \dfrac{\ln S_1 - \ln S_2}{\ln \nu_1 - \ln \nu_2} \pm \dfrac{1}{\ln \nu_1 - \ln \nu_2} \sqrt{\bigg(\frac{\sigma_1}{S_1}\bigg)^2 + \bigg(\frac{\sigma_2}{S_2}\bigg)^2} ,
\end{equation}

\noindent where $S_1$ and $S_2$ are the flux densities at frequencies $\nu_1$ and $\nu_2$, respectively, while $\sigma$ is the corresponding error including flux scale uncertainties. We only display the emission that is above 3 $\times \ rms$ in both frequency bands. The resulting image is shown in the left panel of Fig. \ref{fig:NGC3411_spidx}, while the spectral index error map can be found in Appendix \ref{app:a}. 

\begin{figure}[t!]
\centering
  \includegraphics[scale=0.35]{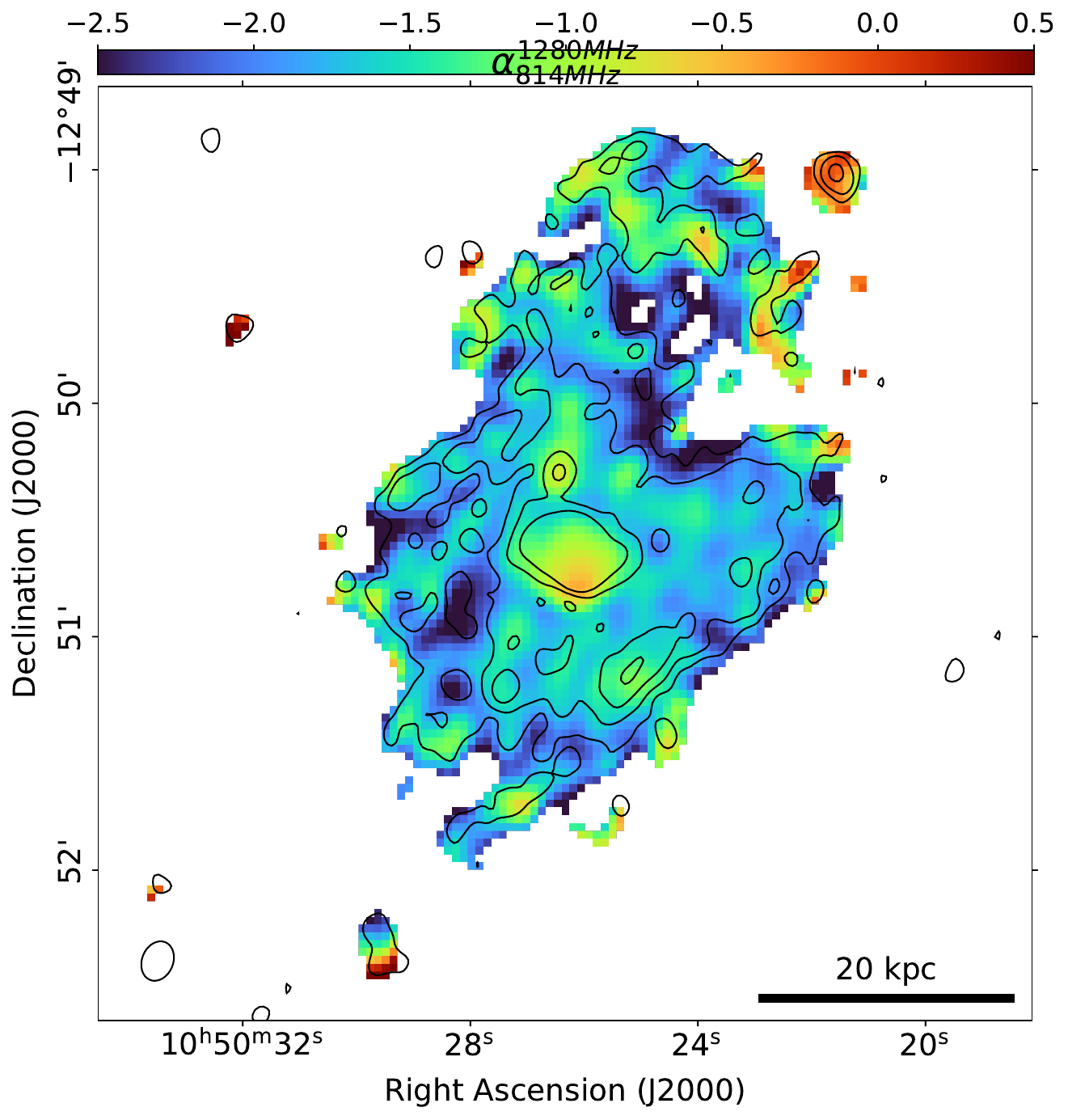}
  \caption{\textit{Left}: Spectral index map of NGC\,3411 between the MeerKAT UHF and L bands. L-band contours are shown in black. \textit{Right}: Spectral index error map of NGC\,3411. Both maps are produced at resolution $10'' \times 8''$, and the beam is shown on the bottom left.}
  \label{fig:NGC3411_spidx}
\end{figure}

The resolved spectral index of the central AGN is consistent with the integrated value reported above, $\alpha_{812 \mathrm{MHz}}^{1280 \mathrm{MHz}} \simeq -0.7$, though the central-most part reaches $\alpha_{812 \mathrm{MHz}}^{1280 \mathrm{MHz}} \simeq +0.2$ due to self-absorption. Interestingly, the resolution is high enough that, within the AGN, we detect a gradient in the spectrum, which is steepening moving north, down to $\alpha_{812 \mathrm{MHz}}^{1280 \mathrm{MHz}} \simeq -1.2$\footnote{We note that we checked for the presence of coordinate offsets in both bands.}. For the diffuse component, we do not detect any clear trend or gradient: the distribution is relatively uniform and shows $<\alpha_{812 \mathrm{MHz}}^{1280 \mathrm{MHz}}> \simeq -1.7$, but the southern and northwestern regions exhibit some very steep values ($\alpha_{812 \mathrm{MHz}}^{1280 \mathrm{MHz}} < -2$). However, uncertainties in these areas are large ($\Delta \alpha \sim0.7$; see the left panel of Fig. \ref{fig:spidxerr}), making these measurements less reliable.

\subsection{NGC\,5903}

\begin{figure*}[t!]
\centering
  \includegraphics[scale=0.40]{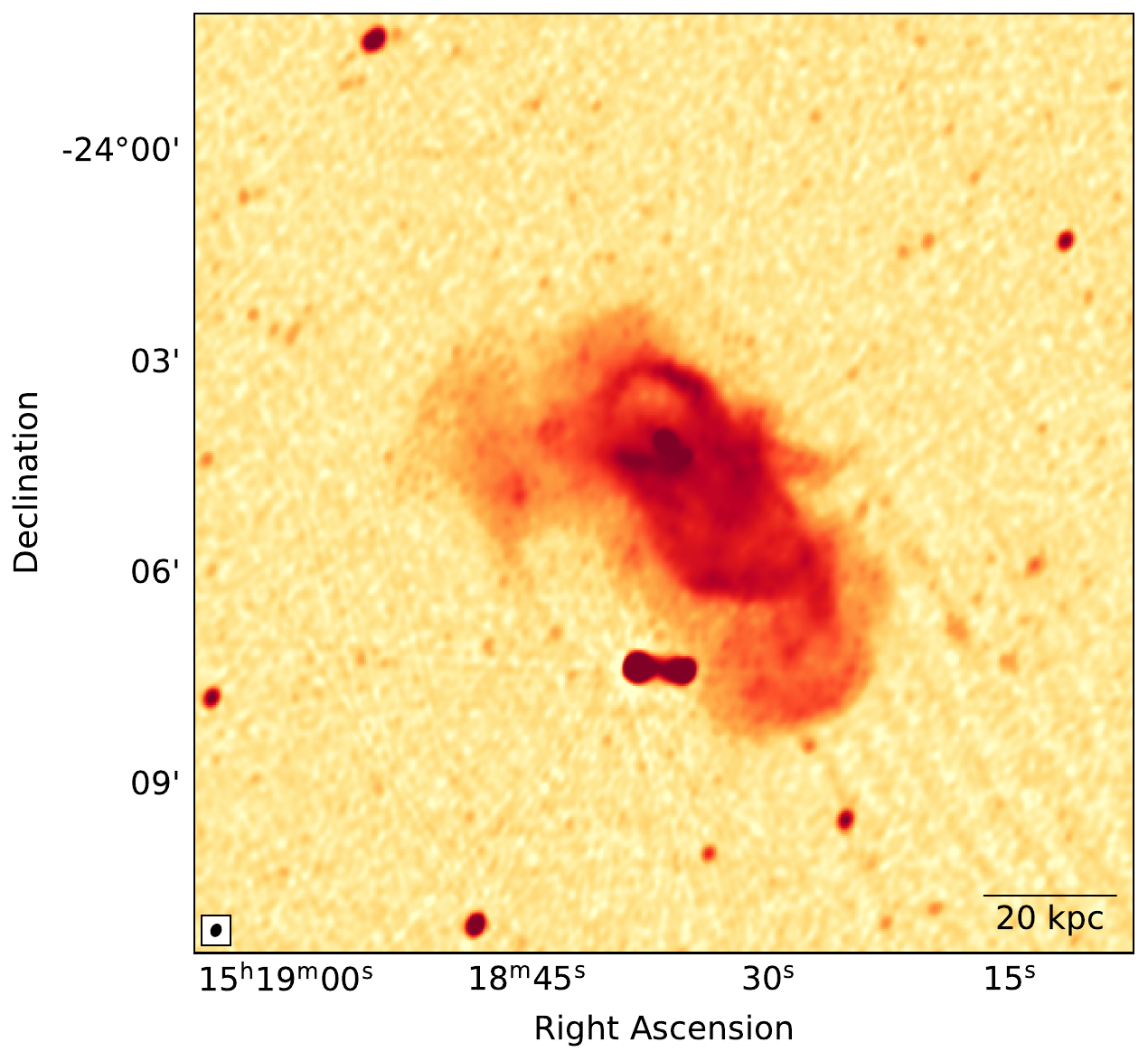}
  \includegraphics[scale=0.40]{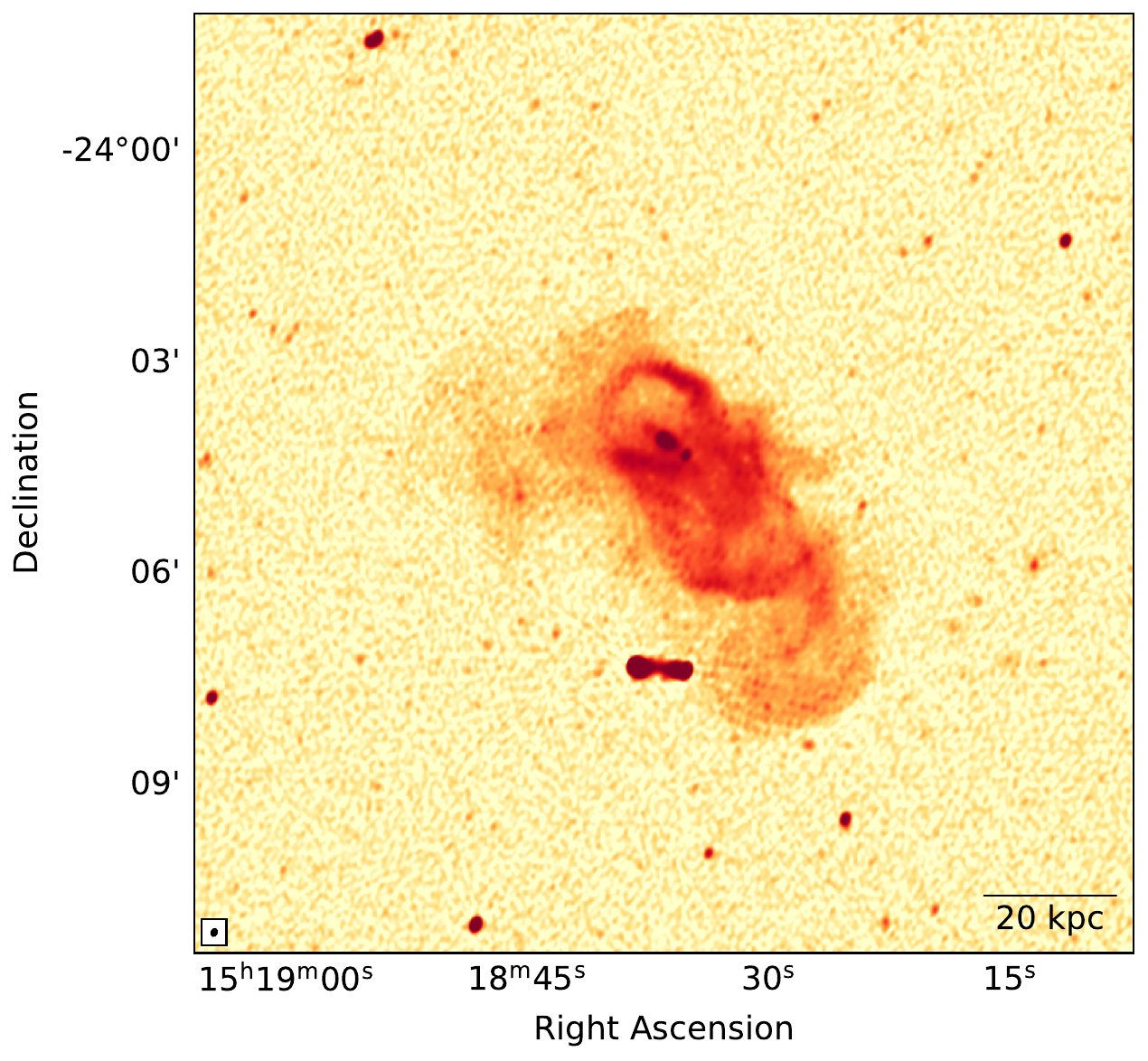}
  \caption{\textit{Left}: UHF-band MeerKAT image of NGC\,5903 produced with {\ttfamily Briggs -0.3}. The rms noise is $\sim$38 $\mu$Jy beam$^{-1}$, and the beam (shown in the bottom left) is $10'' \times 8''$. \textit{Right}: L-band MeerKAT image of NGC\,5903 produced with {\ttfamily Briggs -0.3}. The rms noise is $\sim$19 $\mu$Jy beam$^{-1}$, and the beam (shown in the bottom left) is $6'' \times 5''$.}
  \label{fig:NGC5903}
\end{figure*}

\subsubsection{Overview}

NGC\,5903 is the dominant galaxy of a nearby, low-redshift ($z \sim 0.009$) galaxy group with $\sim$30 galaxy members. No mass estimate is currently available in the literature. This system has been studied in detail, especially in the X-ray band, in \citet{OSullivan_2018}. We summarize here its main properties. This galaxy group exhibits a remarkably complex multiphase intergalactic environment, characterized by the coexistence of cold neutral hydrogen (HI) filaments and a hot IGrM, as well as diffuse radio emission of unclear origin.

A prominent $\sim$110 kpc HI filament crossing NGC\,5903 and a $\sim$75 kpc-wide radio continuum structure overlapping with the same galaxy were identified from VLA observations of the group \citep{Appleton_1990}. The radio structure appears to be partly enclosed by the HI filament, indicating a possible interplay between cold and relativistic plasma components. Combined \textit{Chandra} and \textit{XMM-Newton} X-ray imaging revealed hot ($\sim$0.65 keV) IGrM within the central 150 kpc of the group, as well as a $\sim$35 kpc loop of enhanced X-ray emission extending southwest from NGC\,5903 that encloses the brightest radio emission \citep{OSullivan_2018}. 

The spatial correlation between the X-ray enhancements and portions of the HI filament suggests that dynamics within the group, including tidal interactions and gas accretion, may be shaping its gaseous environment. GMRT spectral measurements of the bright radio emission (see OS18) indicate a steep spectral index, consistent with radiative aging, pointing toward an evolved relativistic plasma likely linked to past activity of the central AGN or to interaction-driven processes. We return to this in the next sections.

\subsubsection{The MeerKAT view}

The left panel of Fig. \ref{fig:NGC5903} shows the UHF-band map of NGC\,5903. We detect compact radio emission that corresponds to the group BGG (the same NGC\,5903). We measure for it a flux density of $S_{814 \rm MHz} = 20 \pm 2$ mJy within 3$\sigma$ contours, which corresponds to $L_{814 \rm MHz} = (3.2 \pm 0.3) \times 10^{21}$ W Hz$^{-1}$ at the group redshift. This source is surrounded by a bright ring of emission, with arcs visible especially to the north and south of the galaxy. We measure a radius of $\sim$8 kpc ($\sim45''$), although it is not clear whether this constitutes a single coherent structure or is instead the projection of multiple filaments that appear physically connected. A complex filamentary structure extends to the south. The whole system is embedded in a fainter, more diffuse component characterized by asymmetrical lobe-like emission to the northeast and south. The northeastern emission appears to be gradually mixed with the surrounding IGrM, while the southern one appears more confined and resembles an inverted mushroom. For the whole structure, excluding the compact AGN, we measure within 3$\sigma$ contours a flux density of $S_{814 \rm MHz} = 515 \pm 52$ mJy, which corresponds to $L_{814 \rm MHz} = (8.4 \pm 0.9) \times 10^{22}$ W Hz$^{-1}$.

In the right panel we show the L-band image. The higher resolution achievable at this frequency enables us to better resolve the inner filamentary structure and the ring-like emission. The inverse mushroom in the south is also clearer, whereas the northeast edge lacks clear morphology. Overall, the system looks very similar to what was detected in the UHF band. For the AGN we measure $S_{1280 \rm MHz} = 15 \pm 2$ mJy, which translates into $L_{1280 \rm MHz} = (2.4 \pm 0.2) \times 10^{21}$ W Hz$^{-1}$. For the diffuse component we estimate instead $S_{1280 \rm MHz} = 325 \pm 33$ mJy within 3$\sigma$ contours, which corresponds to $L_{1280 \rm MHz} = (5.3 \pm 0.5) \times 10^{22}$ W Hz$^{-1}$ at the group redshift. 

We note that these measurements are consistent with those reported in OS18 within the expected uncertainties. We also highlight that the substructures we detect thanks to the high resolution of MeerKAT have not been reported previously in this system. The GMRT data used in OS18 are likely too low in resolution to resolve them, as they only detect a diffuse, amorphous blob. The ring-like structure surrounding the AGN of the BGG, in particular, has never been observed, as it gets easily blended into the diffuse component.

\subsubsection{Spectral index map}

\begin{figure}[h!]
\centering
  \includegraphics[scale=0.34]{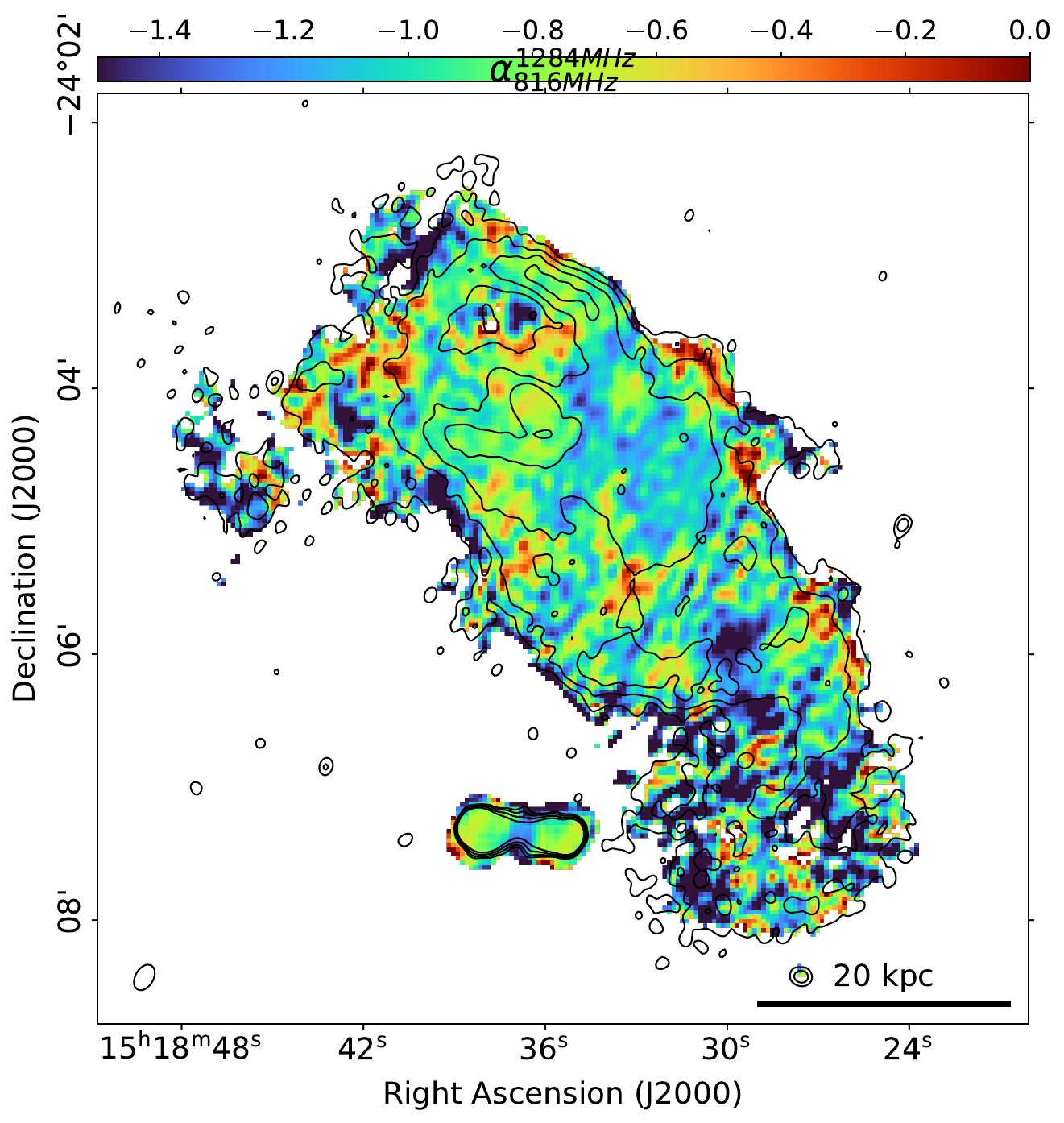}
  \caption{Spectral index map of NGC\,5903 between the MeerKAT UHF and L bands. L-band contours are shown in black.}
  \label{fig:NGC5903_spidx}
\end{figure}

With the luminosity values reported in the previous section, we estimate for the AGN an integrated spectral index of $\alpha_{814 \rm MHz}^{1280 \rm MHz} = -0.61 \pm 0.31$. For the diffuse component we measure instead $\alpha_{814 \rm MHz}^{1280 \rm MHz} = -1.02 \pm 0.31$. These values are confirmed through the spectral index map shown in Fig. \ref{fig:NGC5903_spidx}, which is produced using the same method previously discussed for NGC\,3411. The mean spectral index is $\overline{\alpha}_{812 \mathrm{MHz}}^{1280 \mathrm{MHz}} \sim -1.1$ and is relatively uniform across the central-most part of the extended structure. Flatter values are observed in the AGN region, down to a minimum of $\alpha_{812 \mathrm{MHz}}^{1280 \mathrm{MHz}} \sim -0.2$. Furthermore, we also observe hints of a gradient corresponding to the north arc detected in continuum, where the spectral index flattens moving toward the outskirts of the nonthermal emission down to $\alpha_{812 \mathrm{MHz}}^{1280 \mathrm{MHz}} > -0.5$, akin to radio relics observed in massive galaxy clusters \citep[e.g.,][]{vanWeeren_2019}. However, we note that spectral index errors are significant on the edges of the nonthermal emission, and the limited resolution (especially of the UHF-band map) does not allow us to sample this region with a sufficient number of independent regions, preventing a robust determination of any possible trend. Finally, moving south, the spectral index distribution becomes significantly steeper, reaching $\alpha_{812 \mathrm{MHz}}^{1280 \mathrm{MHz}} \sim -1.5$, although we highlight again that errors in this region are large. We note that a radiative age estimate of $> 350$ Myr was reported in OS18, based on the absence of a spectral break down to 150 MHz. 

\subsubsection{Polarization}

\begin{figure*}[t!]
\centering
  \includegraphics[scale=0.35]{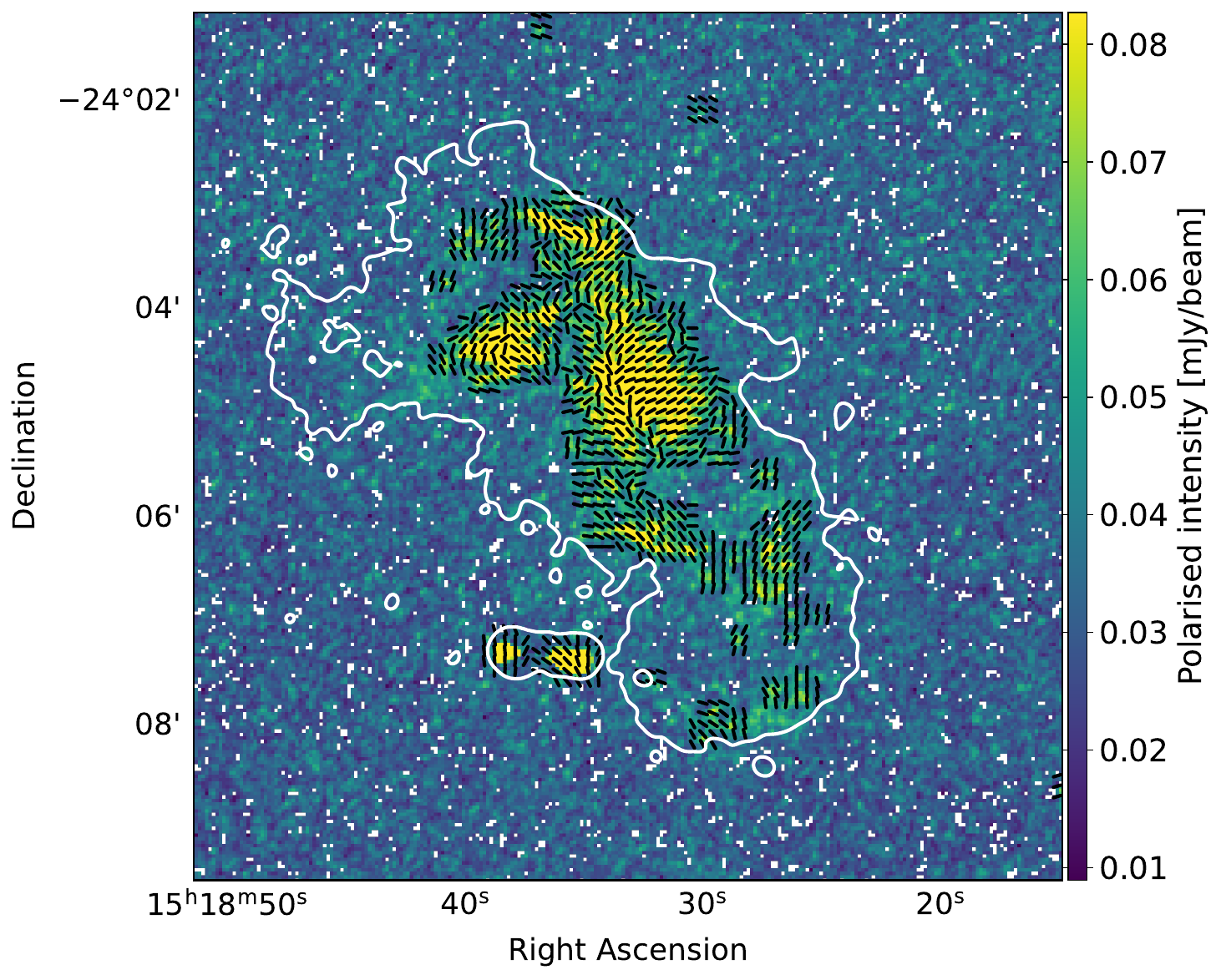}
  \includegraphics[scale=0.35]{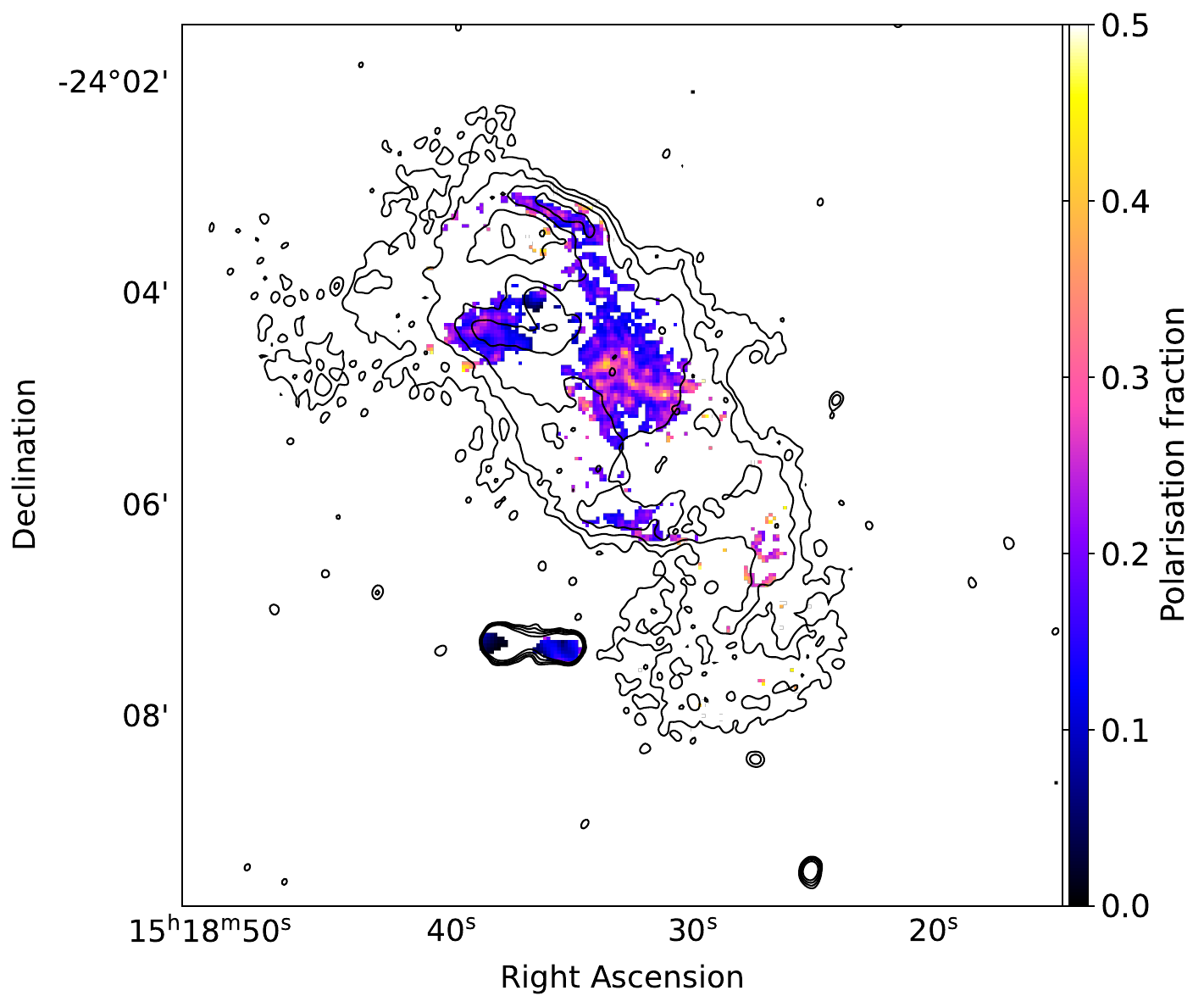}
  \caption{Polarization maps in the L band of NGC\,5903. \textit{Left}: Polarization intensity map (corrected for Ricean bias) with magnetic field vectors (corrected for Faraday rotation) overlaid in black. Total intensity contours are shown in white. \textit{Right}: Polarization fraction map, clipped at 6$\sigma$, with total intensity contours overlaid in black.}
  \label{fig:NGC5903_polarisation}
\end{figure*}

In Fig. \ref{fig:NGC5903_polarisation} we show the polarization intensity map and the polarization fraction map in the L band for NGC\,5903. Polarized emission is clearly detected across the diffuse radio structure observed in continuum, and in particular in the regions of the ring and beyond the AGN, toward the southwest. In general, we observe a fairly structured, filamentary, polarized radio source. The polarization vectors seem to follow the elongated emission and its bends, especially in the southwest, suggesting a coherent, large-scale magnetic field that has been stretched, compressed, or sheared by the group environment. 
The fractional polarization, shown in the right panel, is highest across a line running through the biggest blob of polarized emission, southwest of the AGN. RM and polarization angle maps are available in Appendix \ref{app:b}.

Overall, results from polarization suggest that the environment of NGC\,5903 has been strongly shaped by the group weather: the magnetic field is organized into filaments, the polarization is suppressed in denser or more turbulent regions, and the Faraday rotation shows a complex and spatially varying structure, which may arise from inhomogeneities in the magnetized plasma and/or from a complex magnetic-field configuration.

\begin{figure*}[t!]
\centering
  \includegraphics[scale=0.35]{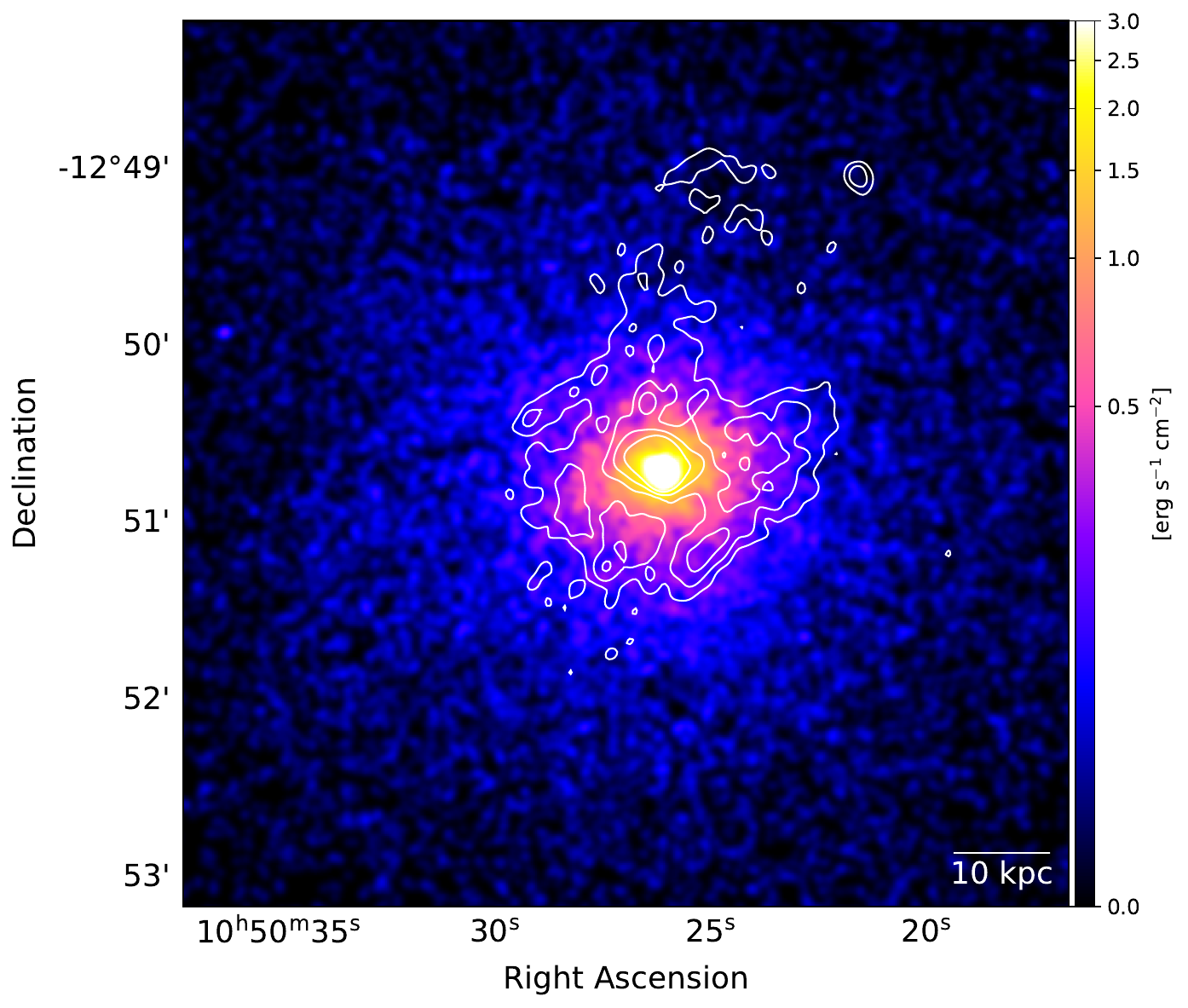}
  \includegraphics[scale=0.39]{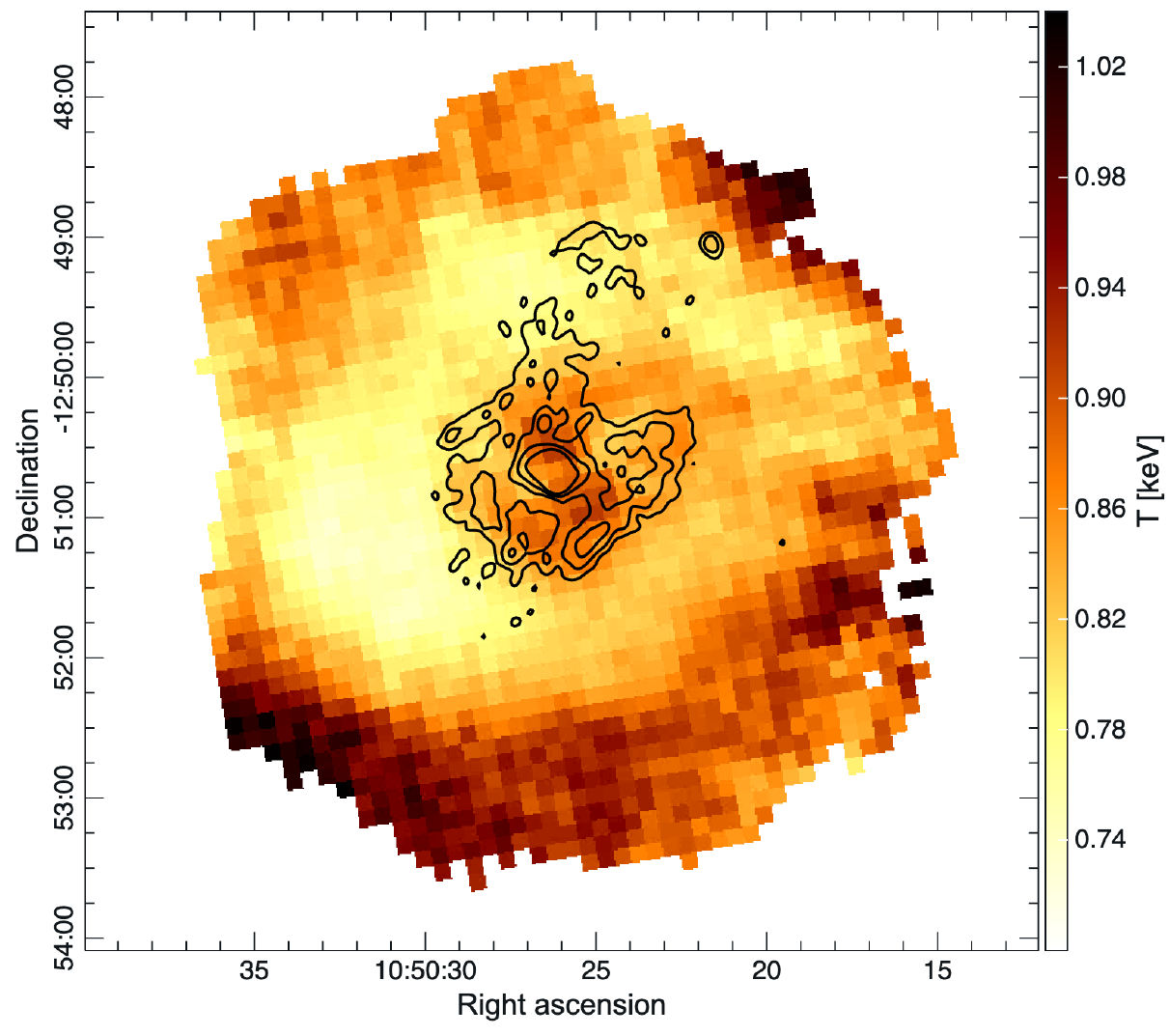}
  \caption{\textit{Left}: 0.3--2 keV \textit{XMM-Newton} exposure-corrected map of NGC\,3411, adapted from OS07. L-band contours are overlaid in white. \textit{Right}: Two-dimensional temperature map of NGC\,3411, with L-band contours, adapted from OS07, overlaid.}
  \label{fig:NGC3411_xray}
\end{figure*}

\section{Discussion}
\label{sec:discussion}

In this section we discuss the results presented in Sect. \ref{sec:results} for our two targets. Our focus will primarily be on the nonthermal emission and its interaction with the IGrM\footnote{We refer to OS07 (for NGC\,3411) and OS18 (for NGC\,5903) for a detailed discussion of their thermal structures and the thermodynamic properties of the hot plasma.}. In general, both systems show signatures consistent with a multiphase group-scale weather cycle in which past AGN outbursts have injected relativistic plasma that is subsequently displaced, stretched, and reprocessed by turbulence, uplift, shear, and bulk IGrM motions. We return to this in the next sections. 

\subsection{NGC\,3411: Group feedback in action?}

NGC\,3411, for which an overlay of our L-band image contours over an archival \textit{XMM-Newton} map is shown in Fig. \ref{fig:NGC3411_xray}, shows highly unusual properties in both thermal and nonthermal emission. The shell of cool ($kT\sim0.7$\,keV) X-ray gas first detected in OS07, which surrounds a hotter ($kT\sim0.84$\,keV) core, is a rare configuration in galaxy groups. Similar cases (e.g., NGC 1407, NGC 4649; \citealt{Humphrey_2006, Kim_2020}) show shallower gradients than the steep rise seen here, while systems with steeper gradients tend to be much more disturbed (e.g., Abell 644; see \citealt{Buote_2005}), unlike the relaxed appearance of NGC\,3411. OS07 discussed two formation scenarios for the cool shell: heating of a formerly cool core by a past AGN outburst, or settling of stripped gas from a member galaxy. They favored the AGN-heating interpretation, as the mass of the shell is too high and no nearby galaxy member is close and bright enough to have provided such a large amount of gas.

Our new MeerKAT L-band observations provide additional, crucial information on the non-thermal component. Compared to the VLA maps used in OS07 (sensitive to structures up to $\sim 10''$ at 1.4\,GHz), MeerKAT reveals much larger-scale emission (major axis $\sim200'' \sim 65$ kpc at 1.4 GHz)\footnote{Note that VLA A-configuration can resolve out large-scale emission; \citet{Giacintucci_2011} reported a major axis of $\sim$80 kpc using GMRT.}. The right panel of Fig. \ref{fig:NGC3411_xray}) shows that the diffuse radio component appears spatially connected to, and in several directions confined by, the cool shell. The MeerKAT image also resolves substructures within the diffuse halo: arc-shaped, edge-brightened filaments, particularly to the west and east, and a clear arc to the north. In the sections below, we investigate three main interpretations for the observed state of this system.

\subsubsection{A mini-halo in a galaxy group?}

The hypothesis of a mini-halo in NGC\,3411 has already been briefly discussed in \citet{Giacintucci_2011}. The observed nonthermal emission surrounds the group BGG, and the irregular, clumpy morphology is quite typical of cluster mini-halos \citep[e.g.,][]{Giacintucci_2014, Giacintucci_2017}. The distribution of the X-ray plasma (see Fig. \ref{fig:NGC3411_xray}) suggests a relatively relaxed (i.e., no recent major merger) environment, akin to cool core clusters where mini-halos are usually detected (see also the review from \citealt{vanWeeren_2019}). 

On the other hand, several counterarguments to the mini-halo hypothesis exist. First, the projected size is atypical, as mini-halos usually extend for $\sim$100-500 kpc, whereas the nonthermal structure in NGC\,3411 extends for $\sim$50 kpc on its major axis. Nevertheless, we also note that there are no reported detections of mini-halos in the galaxy group regime, and their properties are therefore largely uncertain. Another piece of evidence that points against the mini-halo scenario is that OS07 reports no detection of sloshing in the IGrM. In addition, the presence of bright, arc-like, edge-brightened filaments is quite atypical for mini-halos. 

\begin{figure*}[t!]
\centering
  \includegraphics[scale=0.267]{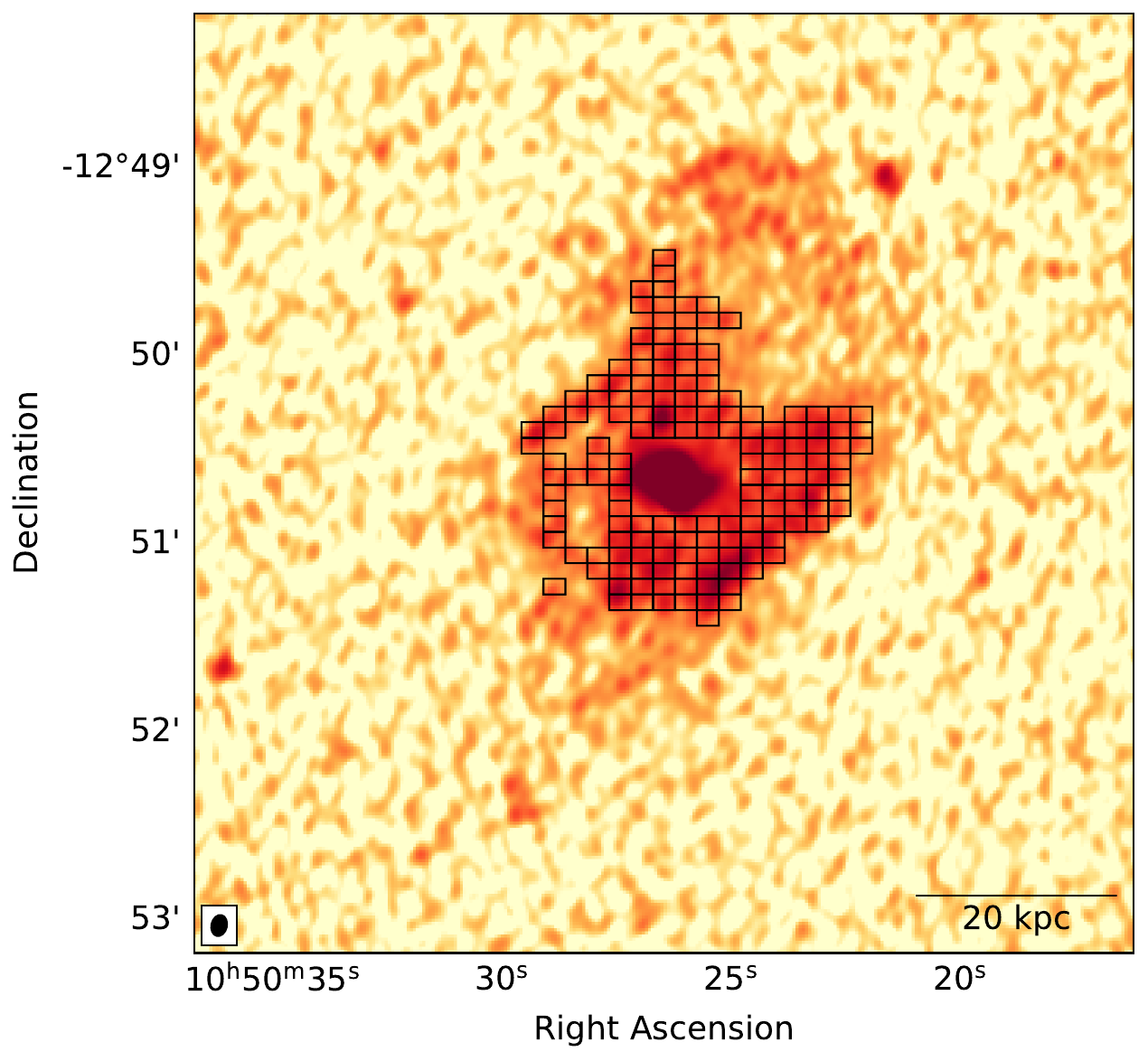}
  \hspace{0.5cm}
  \raisebox{-0.15cm}{\includegraphics[scale=0.232]{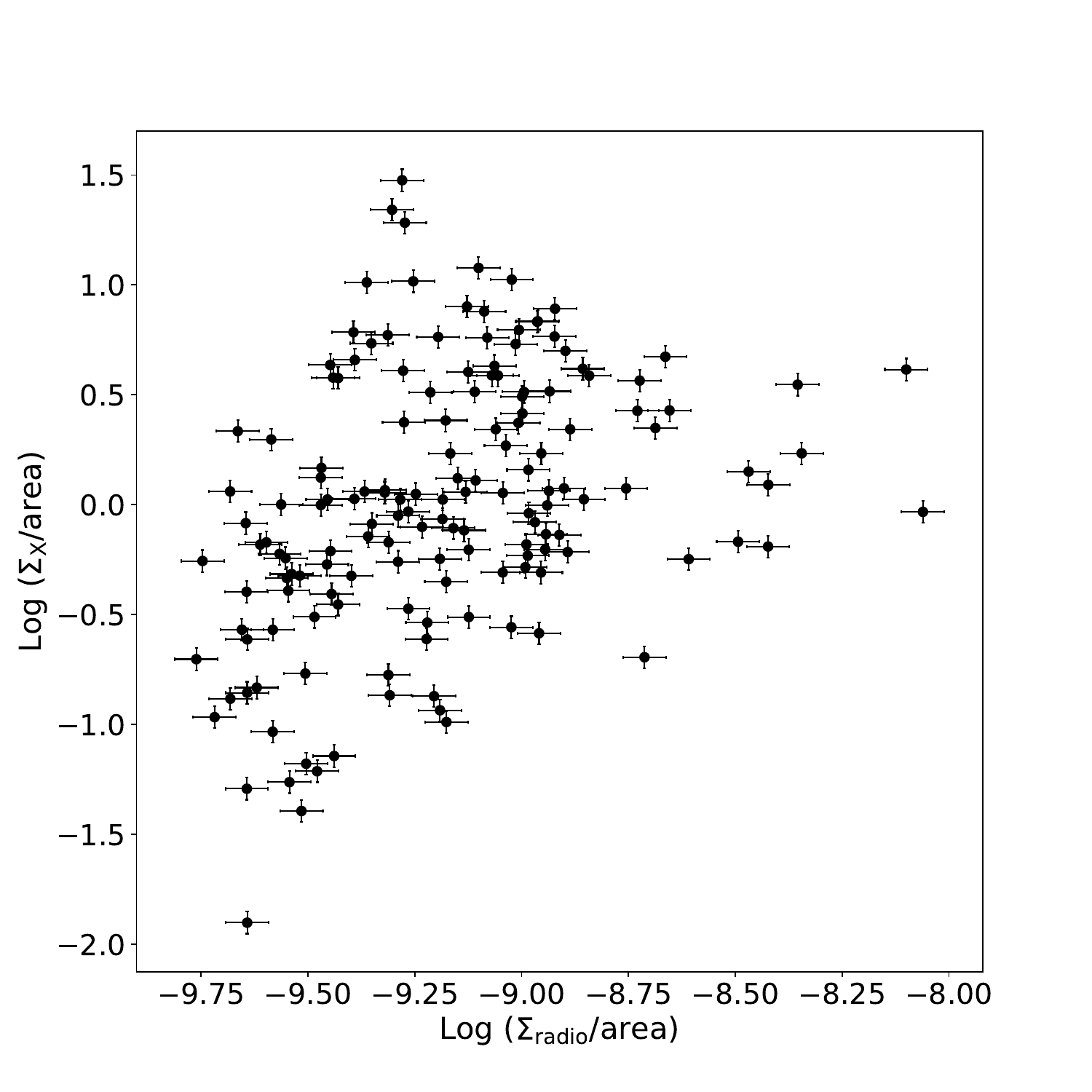}}
  \raisebox{0.1cm}{\includegraphics[scale=0.204]{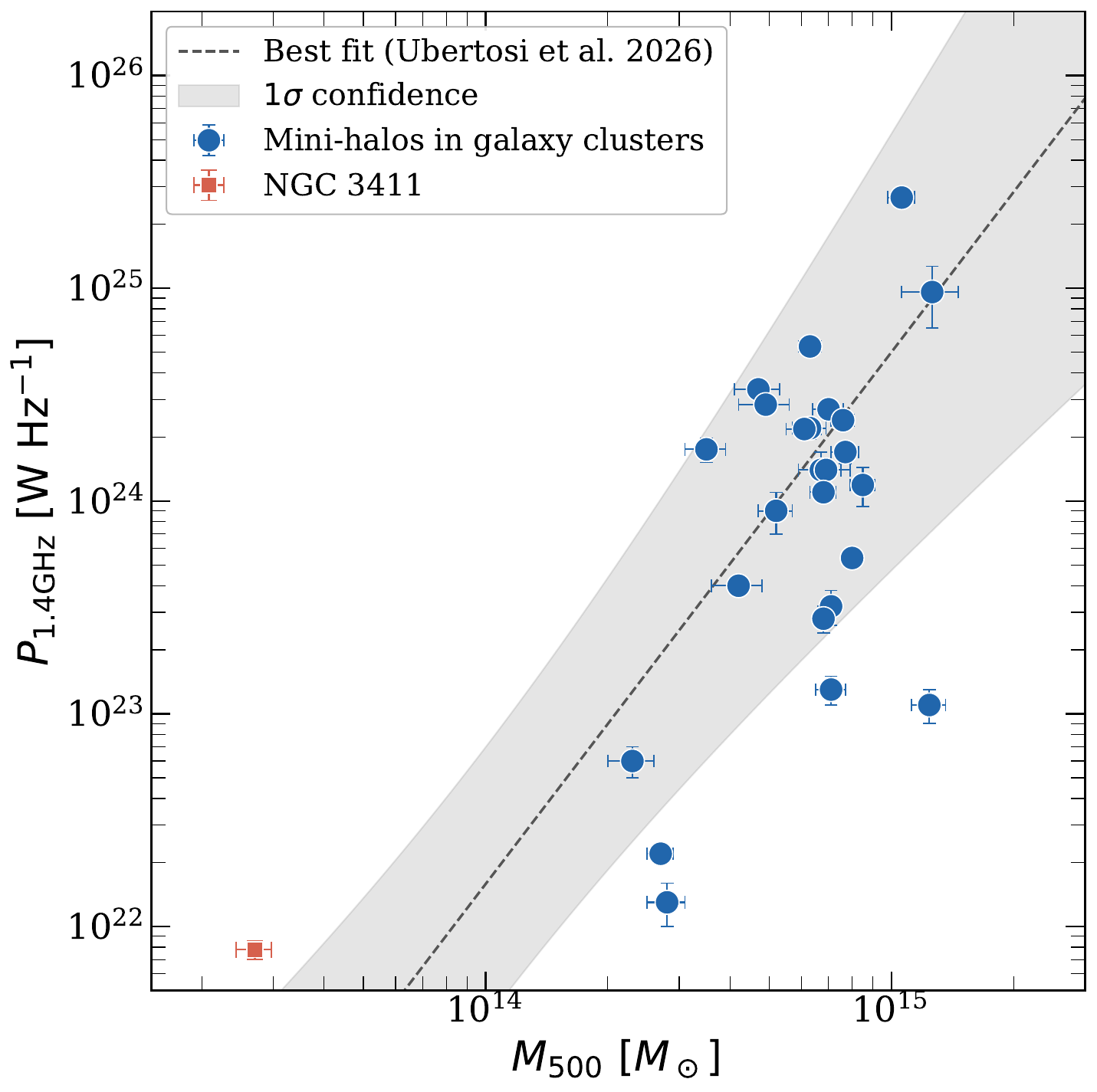}}
  \caption{\textit{Left}: Regions from where the surface brightness in the radio and X-ray bands has been sampled for the point-to-point analysis of NGC\,3411. Each region is as large as the \textit{XMM-Newton} resolution ($15''$). \textit{Center}: Results of the point-to-point analysis. No correlation is found between the radio (x-axis) and the X-ray (y-axis) surface brightness, with a Spearman coefficient of $p \sim 0.45$. \textit{Right}: Mini-halo radio power at 1.4 GHz vs. cluster mass for the galaxy cluster sample studied in \citet{Ubertosi_2026}. The data point for NGC\,3411 is shown in orange. The dashed line shows the best fit from \citet{Ubertosi_2026}, while the gray region represents the $68\%$ confidence band.}
  \label{fig:ptrex}
\end{figure*}

A super-linear correlation is usually observed, in clusters hosting mini-halos, between the X-ray surface brightness of the thermal plasma and the radio surface brightness of the nonthermal emission \citep[e.g.,][]{Gitti_2015, Giacintucci_2019}. We therefore performed a point-to-point analysis by sampling the surface brightness across the extended source in both bands, using the same MeerKAT L-band image of Fig. \ref{fig:NGC3411} and the \textit{XMM-Newton} map between 0.3 and 2 keV. In more detail, the surface brightness was extracted from rectangular regions (see the left panel of Fig. \ref{fig:ptrex}) as large as the beam of the \textit{XMM-Newton} image ($\sim15''$), using {\ttfamily PT-REX} \citep{Ignesti_2022} and masking emission below $3\sigma$, as well as the central AGN. The result, which is shown in the central panel of Fig. \ref{fig:ptrex}, shows that we find no clear correlation between the X-ray and the radio emission, as suggested by the Spearman coefficient ($\rho_S \sim 0.05$) and $p$-value ($p \sim 0.45$). 
A further check is to assess the position of the candidate mini-halo in the known scaling relation between cluster mass and radio power of the diffuse emission \citep[e.g.,][]{Giacintucci_2019}. The result, in the right panel of Fig. \ref{fig:ptrex}, shows that the candidate mini-halo in NGC\,3411 lies $\sim$1 order of magnitude above, in terms of radio power, compared to sources in massive galaxy clusters \citep[e.g.,][]{Ubertosi_2026}. However, it is also worth noting that even a small number of confirmed cluster mini-halos lies outside of the main correlation. In particular, these sources are found to be under-luminous, in the radio band, when compared with the host cluster mass. The properties and mere existence of mini-halos in galaxy groups are largely unknown and therefore unconstrained, as no confirmed detection of this class of sources is currently reported, and the relation is calibrated only upon massive clusters. Nevertheless, the fact that NGC\,3411 lies on the opposite side of the correlation found for galaxy clusters, and that no link is observed between the X-ray and radio surface brightness, represent arguments against the mini-halo hypothesis, albeit with the caveats mentioned above.

\subsubsection{ORCs and radio shocks}

The radio detection of shocks surrounding central galaxies has recently attracted significant attention, particularly following the discovery of ORCs, which seem to be preferentially located in galaxy groups or, more generally, in galaxy overdensities \citep[e.g.,][]{Gupta_2025}. In the case of NGC\,3411, however, we note several important differences. First, the physical scale is substantially smaller because ORCs typically extend over hundreds of kiloparsecs. Moreover, one of the most widely discussed formation scenarios for ORCs involves shocks generated by galaxy–galaxy mergers (see, e.g., the Cloverleaf system discussed in \citealt{Dolag_2023} and \citealt{Bulbul_2024}), whereas NGC\,3411, considered here as the BGG, shows no evidence of recent interactions with other group members or hints of shock fronts. The morphology also differs from that of ORCs, which generally exhibit a relatively regular and spherical structure. Therefore, although these considerations are not conclusive, the diffuse radio component associated with NGC\,3411 might represent a different class of source and, given the currently available data and ORC formation models \citep[e.g.,][]{Bulbul_2024}, cannot be securely classified as an ORC. However, it is also worth noting that we are still in the early phases of the study of these sources: detailed IGrM temperature maps or resolved spectral index images are not available for any of the confirmed ORCs, and it is therefore currently not possible to do a proper comparison with the environment and spectral properties of NGC\,3411. 

While the observed structure might not be a typical ORC, shocks may still have played an important role in shaping the current state of this galaxy group. Indeed, the contribution of shocks in reheating the cooling plasma of clusters and groups has been known for some years now. Cluster-scale shocks can be induced in massive systems due to AGN jets and lobes expanding through the ICM, with some noteworthy examples being Hydra A \citep{Nulsen_2005}, M87 \citep{Forman_2005}, and 3C 444 \citep{Croston_2011}. Large shocks have also been detected in galaxy groups \citep[e.g.,][]{Liu_2019}. In all these systems, the shock contribution to heating is always found to be of the same order as mechanical heating from jet-induced cavities. However, the case of NGC\,3411 looks somewhat different. First of all, there is no hint of cavities in the thermal plasma, and indeed, no jets or lobes are detected in the radio band, not even with the new MeerKAT data. Furthermore, we are at a galactic scale: for the west candidate shock, for example, we measure a total projected extent of $\sim$30 kpc. 

We can still try to assess the potential role of a shock in raising the central core temperature from $kT \sim 0.7$ keV to $kT \sim 0.85$ keV, corresponding to the values reported by OS07 for the cool shell and the core, respectively. For these calculations, we adopted the thermodynamic properties of the IGrM given in that work. As estimated in their Sect. 4.1, the energy required to heat the core by $\Delta T \sim 0.15$ keV is $E_{\rm heat} \sim 2 \times 10^{57}$ erg. OS07 also report that, in the case of mechanical heating from the central AGN, the energy injection rate would be $L_{\rm mech} \sim 8.2 \times 10^{41}$ erg s$^{-1}$, slightly lower than the radiative cooling losses of $L_{\rm cool} \sim 1.13 \times 10^{42}$ erg s$^{-1}$ in the core. Because no shocks were detected in their data, OS07 suggested that a relatively recent, powerful AGN outburst might have supplied the energy needed to heat the group core to its current state. 

However, the arcs identified through our new MeerKAT observations raise the possibility that, if they are signatures of shocks, they could have played a role in reheating the group core during a past period of AGN activity. Indeed, the very steep radio spectrum ($\bar{\alpha} \sim -1.7$ between the MeerKAT UHF- and L-band images) is consistent with heavily aged synchrotron plasma from an old AGN outburst. With an injection spectral index $\alpha_{\rm inj} \sim -0.6$\,--\,$-0.7$ typical of AGN jets \citep[e.g.,][]{Murgia_1999}, the observed steepening of $\Delta\alpha \sim 1.0$--$1.1$ implies substantial radiative aging, pointing to a plasma age of at least several tens to $\sim$100\,Myr depending on the assumed magnetic field strength. We note that in this scenario --- an AGN-driven expanding cocoon shock --- the spectral index of the radio plasma reflects the age and history of the jet plasma, rather than the properties of the shock front itself. Consequently, it is not appropriate to derive the shock Mach number from the radio spectral index similarly to what is done for diffusive shock acceleration, such as for radio relics. Instead, we estimated the Mach number from the X-ray thermodynamic data of OS07. The temperature ratio between the group core and the cool shell, $kT_{\rm core}/kT_{\rm shell} = 0.85/0.7 \approx 1.21$, provides a direct constraint on the post- to pre-shock temperature jump. Inverting the Rankine--Hugoniot condition for $\gamma = 5/3$, 
\begin{equation}
\frac{T_2}{T_1} = \frac{(5M^2 - 1)(M^2 + 3)}{16\,M^2},
\label{eq:RH_T}
\end{equation}
a ratio of $T_2/T_1 = 1.21$ yields $M \approx 1.2$. 

The energy associated with the shock \citep{Liu_2019} can be estimated as
\begin{equation}
E_{\rm shock} \sim P_1\,V\left(\frac{P_2}{P_1} - 1\right) = \Delta P \cdot V.
\label{eq:Eshock}
\end{equation}

\noindent Here, the pressure values are derived directly from the X-ray thermodynamic profiles of OS07. For a ionized intra-group plasma, the thermal pressure can be estimated as $P \approx 1.9n_e k_B T $ for a plasma with solar abundances, where $n_e$ is the electron density and $T$ is the gas temperature. In the analysis of OS07, the relevant thermal state of the gas is characterized by $T_1 \approx 0.7~{\rm keV}$ in the cool shell (pre-shock), and $T_2 \approx 0.85~{\rm keV}$ in the core (post-shock). The corresponding electron densities are taken from the deprojected X-ray density profiles of OS07. The volume of the heated gas is $V = \frac{4}{3}\pi R^3$, with $R = 20$ kpc. The shock energy is then estimated to be $E_{\rm shock} \sim 5.45 \times 10^{58}$ erg, approximately $30\times E_{\rm heat}$. We then calculated the shock age as $t_{\rm age} \sim d/v_{\rm shock}$, where $d \sim 20$ kpc is the projected distance to the candidate shock. and $v_{\rm shock} = M \times c_s$. The sound speed in the pre-shock gas is
\begin{equation}
c_s = \sqrt{\frac{\gamma\,k_B T}{\mu\,m_H}} \approx 430\ \mathrm{km\,s}^{-1}
\label{eq:cs}
\end{equation}
for a $T$ of $\sim 0.7$\,keV. With $M \approx 1.2$ we obtain $v_{\rm shock} \approx 516$\,km\,s$^{-1}$ and $t_{\rm age} \approx 38$\,Myr. We therefore estimated a shock power:
\begin{equation}
L_{\rm shock} = \frac{E_{\rm shock}}{t_{\rm age}} \approx 5 \times 10^{43} \mathrm{erg\,s}^{-1}.
\label{eq:Lshock}
\end{equation}
Thus, based on these assumptions, $L_{\rm shock}$ exceeds $L_{\rm cool}$ by more than one order of magnitude. The energy available from the AGN-driven shock is therefore not only sufficient to balance cooling, but would easily provide the $E_{\rm heat} \sim 2 \times 10^{57}$\,erg required to reheat the core, even if only a modest fraction of the shock energy couples thermally to the IGrM. The picture that would emerge is one in which a past AGN outburst drove an expanding cocoon shock into the group IGrM. The shock has since propagated to $\sim$20--30\,kpc, leaving behind heavily aged radio plasma visible in the MeerKAT images and a reheated core. We note that the very steep spectrum ($\alpha \sim -1.7$) implies a synchrotron age potentially longer than the dynamical shock age estimated above; this could indicate that the radio plasma was energized during an even earlier episode of AGN activity, or that the local magnetic field is weaker than the equipartition value, both of which are plausible in a fading outburst scenario. It is also important to note that OS07 assumed $\Delta T = 0.15$\,keV based on the cool shell temperature ($kT = 0.7$\,keV). If the core were initially cooler than this value, as is plausible for a relaxed cooling core prior to the outburst, additional energy would be required; this could plausibly be supplied by shocks, based on our estimates.

Nevertheless, the above estimate should be regarded as an approximation, since it assumes that the core-to-shell temperature contrast is the projected signature of a shock discontinuity. For $M \approx 1.2$, the Rankine--Hugoniot conditions predict a density jump of $\sim$1.3 and an X-ray surface brightness jump of $\sim$1.7. Projection effects and the finite \textit{XMM-Newton} spatial resolution could further dilute such a weak edge. In addition, no X-ray cavities are detected at any scale, despite these being a known signature of AGN-driven shocks in groups and clusters \citep[e.g.,][]{Nulsen_2005, Forman_2005}. Thus, while the absence of a clear discontinuity does not strictly exclude a weak disturbance, the lack of an X-ray edge or cavity prevents the radio arcs from being interpreted as evidence of a strong AGN-driven shock.

Given this evidence, we argue that the arc-like filaments may trace fossil AGN plasma that has been advected, stretched, and weakly compressed along the thermodynamic boundary of the cool shell, with shear and magnetic draping locally amplifying the magnetic field \citep{Dursi_2008, DeGasperin_2017, Brienza_2022}. Such circulation is naturally expected within a self-regulated chaotic cold accretion cycle, where subsonic turbulence, uplift, weak compressions, and recurrent mechanical feedback continuously reshape the multiphase atmosphere \citep{Gaspari_2020, Wittor_2020, Cammelli_2026}. These processes may gently reenergize aged synchrotron plasma without requiring the temperature and pressure jump of a strong shock, while remaining consistent with the relaxed large-scale X-ray morphology of NGC\,3411.

\subsubsection{A line-of-sight radio galaxy?}

A third, more qualitative possibility is that we might be observing the radio galaxy as closely aligned along the line of sight. In this case, the bright nuclear component would correspond to the Doppler-boosted core, the diffuse halo to a foreground overlap of the radio lobes, and the absence of jets even at MeerKAT resolution would be naturally explained by the small inclination angle. The relatively uniform spectral index across the source and the positive value observed instead in the very center of the AGN core are also qualitatively consistent with a pair of evolved, aged lobes observed in projection (e.g., \citealt{Hardcastle_2020}). In this scenario, the observed confinement of the source within a hotter halo, surrounded by the cool shell, would be easily explained if the core had been reheated by the energy deposited through the expanding radio jets. It is also worth mentioning that the non-detection of polarized emission corresponding to the extended source might be relevant. Radio lobes, even when aged and/or remnant, are typically detected with fractional polarization of $\sim 10\%-30\%$ at gigahertz frequencies \citep[e.g.,][]{Hardcastle_2020}. However, if the extended emission is embedded within the IGrM, internal Faraday dispersion and beam depolarization along the full path length through the magnetized IGrM might offer a natural explanation for the non-detection of polarization \citep[e.g.,][]{Burn_1966, Sokoloff_1998}, especially given the relatively short exposure time. 
This interpretation remains necessarily speculative, although plausible. It avoids some of the difficulties of the mini-halo and strong-shock scenarios, but the present data do not establish it as preferred over the other hypotheses discussed above. In particular, the lack of direct constraints on the three-dimensional geometry and orientation of the radio emission prevents further, accurate assessments about the source energetics, age, and interaction with the surrounding IGrM. In the future, follow-up observations at extremely high resolution (e.g., very long baseline interferometry) could potentially provide useful information about jet orientation, and thus might be able to shine more light on this.

\subsection{NGC\,5903: Group weather and weak shocks}

\begin{figure*}[t!]
\centering
  \includegraphics[scale=0.25]{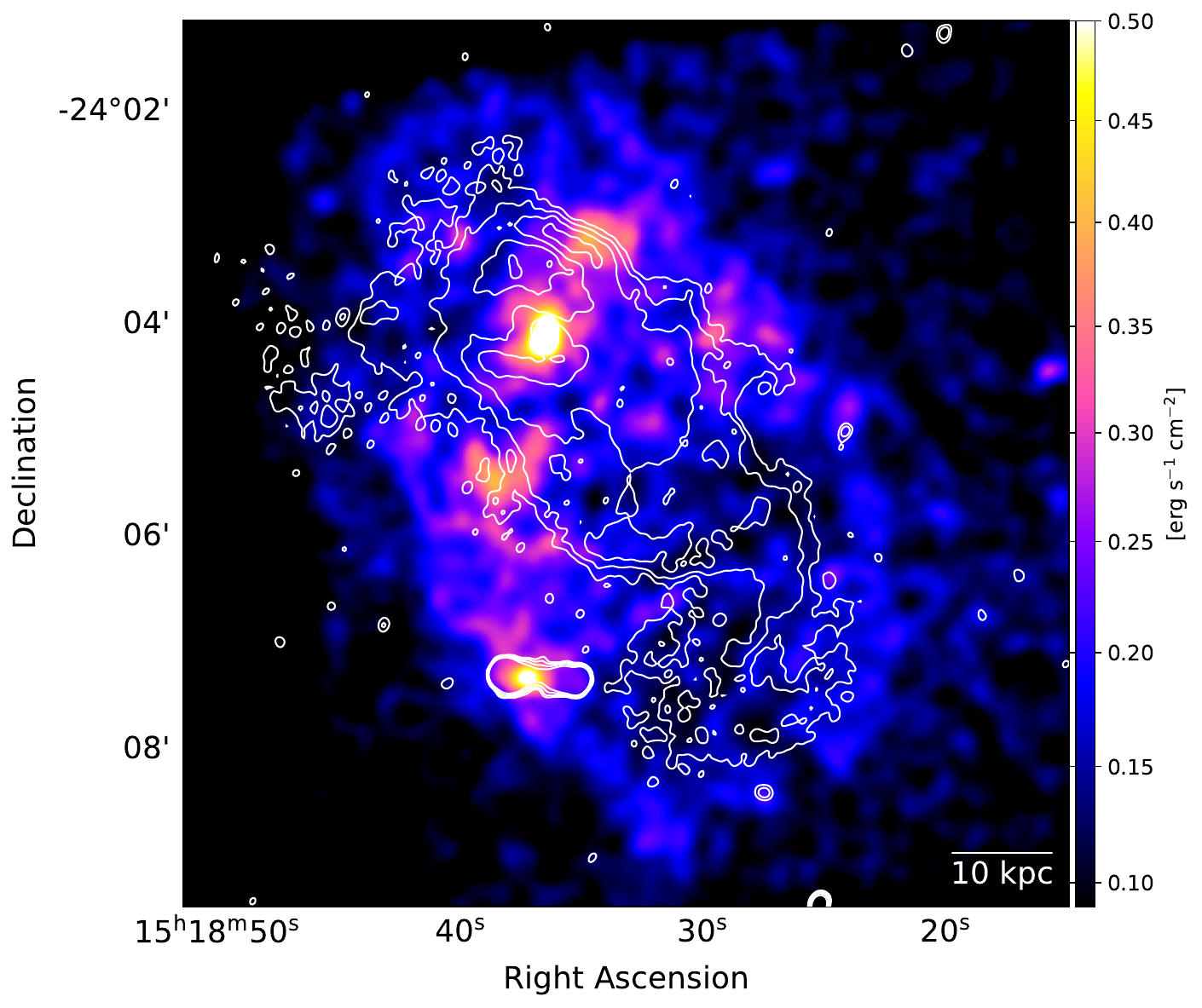}
  \includegraphics[scale=0.25]{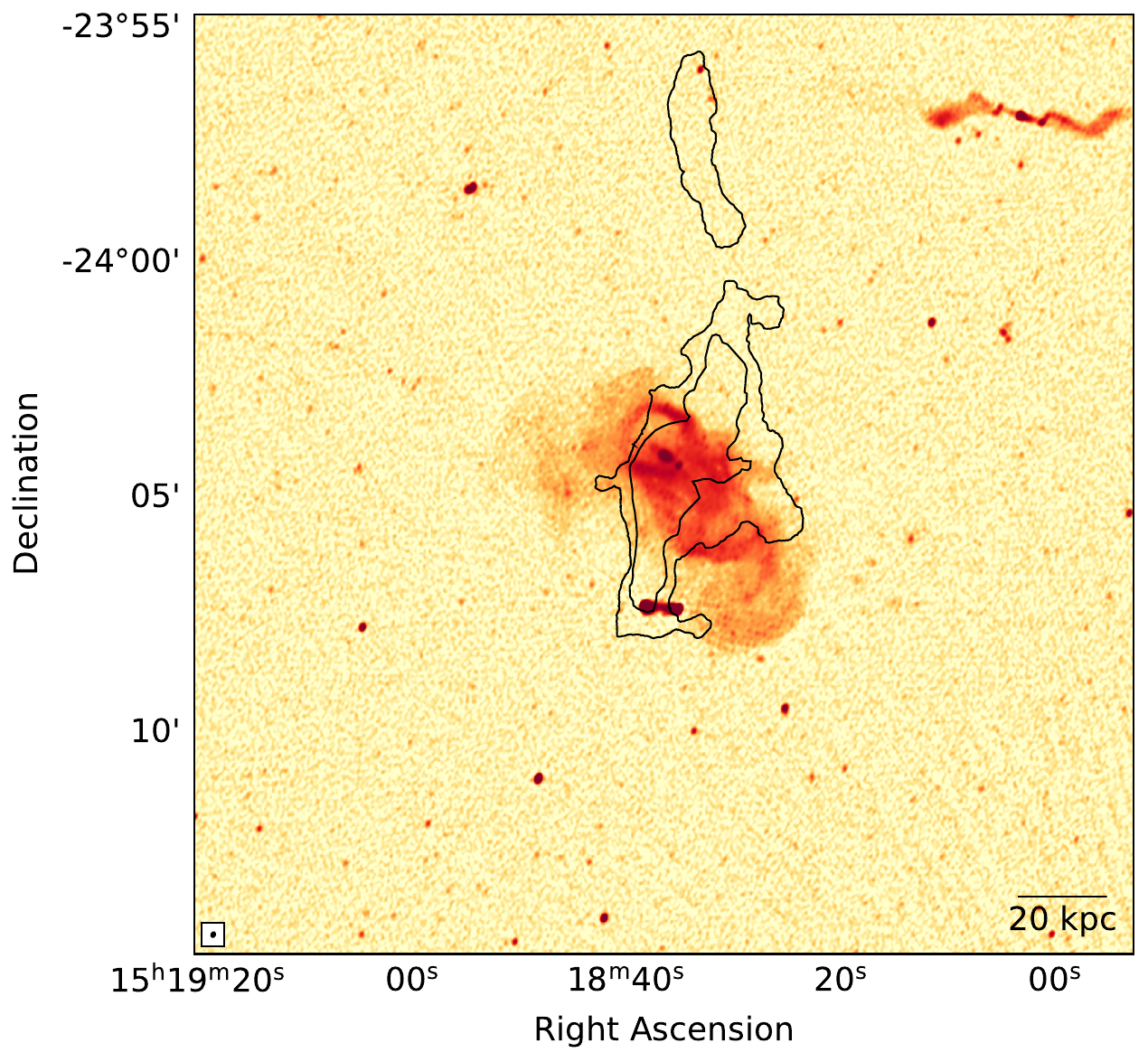}
  \includegraphics[scale=0.265]{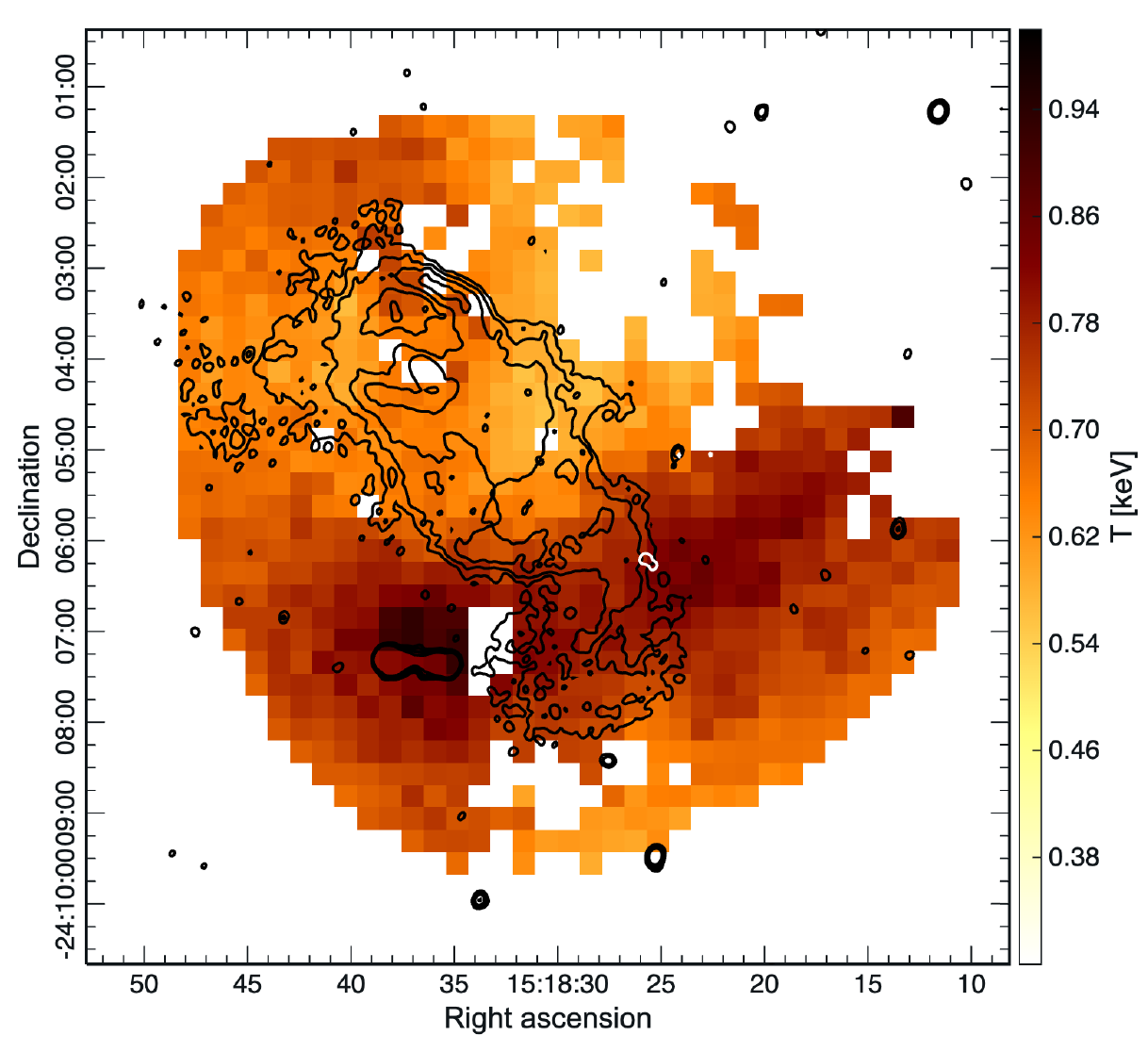}
  \caption{\textit{Left}: 0.5-2 keV \textit{Chandra} exposure-corrected map of NGC\,5903, adapted from OS18. L-band contours are overlaid in white. \textit{Center}: L-band image of NGC\,5903 (same as Fig. \ref{fig:NGC5903}) with HI contours from OS18, originally derived from \citet{Appleton_1990}, overlaid. \textit{Right}: Two-dimensional temperature map of NGC\,5903, with L-band contours, adapted from OS18, overlaid.}
  \label{fig:NGC5903_xray}
\end{figure*}

As extensively discussed in OS18, the case of NGC\,5903 is at least as complex as that of NGC\,3411. The radio morphology is highly unusual: the AGN and the ring-like structure are embedded within a much larger diffuse halo extending predominantly along the east and southwest directions. This overall configuration strongly suggests a significant interaction with the surrounding environment, similar to what has been reported in other galaxy groups such as Nest200047 \citep{Brienza_2021} and NGC\,507 \citep{Brienza_2022}. In the left panel of Fig. \ref{fig:NGC5903_xray}, we show the MeerKAT L-band contours overlaid on the same \textit{Chandra} image used by OS18. As already noted in that work, the X-ray emission is strongly disturbed and asymmetric, with the IGrM distributed along an elongated structure roughly aligned with the radio source. The surface brightness peaks at the AGN position but shows several secondary enhancements and depressions, indicating substantial substructure in the thermal gas. The X-ray emission is not smooth, suggesting ongoing dynamical activity and/or strong interaction with the AGN. The brightest X-ray regions spatially correlate with the inner radio features, including the ring, while more extended, lower-surface-brightness X-ray emission surrounds the outer radio contours, consistent with a strong coupling between the relativistic plasma and the surrounding hot atmosphere. The southwest X-ray depression has indeed been identified as a cavity in OS18, who also inferred a jet power of $P{\rm cav} \sim 8.1 \times 10^{40}$ erg s$^{-1}$ required to inflate it.

OS18 considered two main scenarios to explain the current state of the system: an interaction-triggered AGN outburst with sufficient enthalpy to shape the IGrM, and a high-velocity collision between a galaxy and the H\,I filament. However, neither mechanism alone can account for all of the observed properties. They therefore proposed a combined scenario in which a tidal interaction with an infalling galaxy (ESO 514-3) first produces the H\,I emission (see the central panel of Fig. \ref{fig:NGC5903_xray}) and triggers an AGN outburst in NGC\,5903, inflating radio lobes and generating the observed X-ray loop through uplift and/or compression. Subsequently, a high-velocity passage of another group member (most plausibly NGC\,5898) shocks parts of the filament and the surrounding diffuse H\,I, producing enhanced temperatures in the southern X-ray loop and the low-surface-brightness, steep-spectrum radio emission (see the right panel of Fig. \ref{fig:NGC5903_xray}). In this framework, the AGN outburst is the primary driver of the bright radio/X-ray structure, but collisional shock heating and interactions with the H\,I filament also play an important role in shaping the present-day group environment. OS18 also noted, however, that their data did not reveal an unambiguous shock or a clear spectral gradient indicative of shock-driven acceleration.

The MeerKAT observations presented here provide crucial new information. The arc-shaped filaments that form the newly detected ring-like structure in the continuum maps (Fig. \ref{fig:NGC5903}) have no obvious optical or IR counterpart and surround the BGG at projected distances of $\sim$10 kpc in the northern sector and $\sim$2 kpc in the southern sector. The spectral-index map at the location of these features shows significant flattening that closely follows the curved morphology of the arcs (Fig. \ref{fig:NGC5903_spidx}). This behavior, which was not evident in OS18 likely due to the lower resolution, is difficult to reconcile with a purely passively aging population of relativistic electrons and instead points to recent particle (re-)acceleration and/or adiabatic compression of preexisting radio plasma. The polarization maps further reveal a relatively ordered magnetic-field structure in this region, especially along the north arc, supporting the idea that the synchrotron-emitting plasma has been dynamically compressed or sheared.

Taken together, these properties qualitatively resemble relic-like features in cluster environments, where shocks can re-accelerate seed cosmic-ray electrons through diffusive shock acceleration (DSA) and simultaneously amplify magnetic fields \citep[e.g.,][]{ensslin_1998, vanWeeren_2019}. However, the north arc in NGC\,5903 is only $\sim$20 kpc in size and lies well within the central region of a galaxy group. Classical radio relics are typically much larger and are associated with merger shocks propagating through the ICM. Any disturbance relevant here must therefore be a much weaker and more localized compression front, likely generated by galaxy-galaxy interactions, minor merger activity, or group-weather motions, rather than by a group-scale merger event. Similar interaction-driven shocks and compressive disturbances are known to occur in dynamically active group environments; the clearest example is Stephan's Quintet, where a high-velocity galaxy collision has produced shock-driven features detected from X-ray to molecular gas \citep[e.g.][]{machacek_2004, appleton_2006, guillard_2009, cluver_2010, osullivan_2009}. Although the physical scale and energetics of NGC\,5903 are much smaller than in Stephan's Quintet, such systems demonstrate that shock- or compression-driven re-acceleration of relativistic plasma is not restricted to massive clusters. The arcs in NGC\,5903 may trace weak shocks or compression fronts acting on an older population of synchrotron-emitting electrons, thereby flattening the local spectrum and enhancing the polarized emission.

Nevertheless, the lack of a direct thermodynamic shock measurement in the X-ray data prevents a firm identification of the arcs as genuine diffusive shock acceleration relics. A more conservative interpretation is therefore that they trace weak compression and/or mild re-acceleration of fossil plasma in the group core. This scenario naturally explains their small physical size, curved morphology, local spectral flattening, and relatively ordered magnetic-field structure while remaining fully consistent with the disturbed dynamical state of NGC\,5903. Overall, our observations are consistent with the picture proposed by OS18, while the multifrequency coverage and higher angular resolution of MeerKAT allow us to place stronger constraints on the physical interpretation. The large-scale radio emission permeating the group is most plausibly the remnant of a past AGN outburst from NGC\,5903, likely fueled, at least in part, by H\,I gas stripped during tidal interactions within the system, likely involving NGC\,5898. In this picture, the extended lobes were deposited during an earlier phase of activity and subsequently underwent passive radiative aging, which naturally explains the steep spectral indices measured over much of the diffuse emission (Fig. \ref{fig:NGC5903_spidx}).
Within this framework, subsequent group weather, bulk motions, and buoyant uplift likely stretched and displaced the aged radio plasma, helping to form the southern extension and excavate the X-ray cavity visible in the left panel of Fig. \ref{fig:NGC5903_xray}. The steep spectral indices measured in the southern and northeastern extensions are consistent with old, radiatively aged synchrotron plasma, whereas the flatter spectrum observed in the central region, particularly along the ring-like arcs, indicates that a second mechanism must be operating there. The occurrence of mild re-acceleration and/or adiabatic compression of fossil electrons, accompanied by magnetic-field amplification, is further supported by the polarization properties of the arcs, which suggest a relatively ordered field configuration rather than a purely chaotic remnant structure. We therefore conclude that the radio source in NGC\,5903 is best understood not as a single, uniformly aging lobe system, but as fossil AGN plasma that has subsequently been redistributed and locally reenergized by the disturbed dynamics of the group. In this sense, the ring-like arcs may represent the small-scale signature of interaction-driven compression or weak shocks in the group core, rather than classical radio relics in the cluster sense. This scenario simultaneously explains the old age of the large-scale diffuse emission, the local spectral flattening and enhanced polarization of the arcs, the correspondence with the disturbed X-ray morphology, and the close connection between the radio plasma and the tidally disrupted gaseous environment.

\section{Conclusions}
\label{sec:conclusions}

We have presented new MeerKAT UHF- and L-band observations of the galaxy groups NGC\,3411 and NGC\,5903, complemented with archival \textit{Chandra} and \textit{XMM-Newton} data. These new observations, which are part of a pilot project that has observed a total of 16 systems, have revealed previously undetected nonthermal substructures in both galaxy groups. Our main results can be summarized as follows:

\begin{itemize}

\item In NGC\,3411, extended radio emission surrounds the central AGN up to $\sim$70 kpc distance, and includes several arc-like structures on scales of $\sim$20-30 kpc. We detect a relatively steep and uniform spectral index $\alpha_{814 \rm MHz}^{1280 \rm MHz} \sim -1.7$ across the radio source. X-ray maps indicate a rather relaxed system.\\

\item While the morphology of the source is akin to cluster mini-halos, we show that the absence of a clear radio--X-ray surface-brightness correlation currently disfavors a standard mini-halo interpretation, although cluster mini-halo scaling relations remain poorly constrained in the group regime. Available X-ray data provide no direct evidence for a strong shock in NGC\,3411. The arcs may instead trace aged AGN plasma reprocessed by weak compression, shear, and magnetic draping along the cool-shell interface. A line-of-sight radio-galaxy geometry also represents a possible explanation, though it is not trivial to test this hypothesis with the current data.\\

\item In NGC\,5903, which shows strong indications of disturbances from X-ray images, the AGN is embedded within a much larger nonthermal structure that extends for $\sim$100 kpc in the northeast--southwest direction. This emission is composed of a number of  substructures, as it shows a thin ring around the AGN and multiple filaments. The mean spectral index is $\alpha_{816 \rm MHz}^{1284 \rm MHz} \sim -1$, but it varies across the source, reaching very steep values $\alpha_{816 \rm MHz}^{1284 \rm MHz} < -1.5$ in the southern region. Additionally, we detect polarized emission that is especially prominent around the central AGN, with magnetic field vectors closely tracing the overall structure. The ring also exhibits a relatively ordered polarization structure compared to the surrounding plasma.\\

\item The features observed in NGC\,5903 are indicative of a system strongly affected by the interplay between past AGN outbursts and group weather. Fossil plasma injected by the AGN has likely been stretched and redistributed across the group environment. The observed properties, further supported by our polarization data, point toward mild re-acceleration or adiabatic compression of fossil electrons.

\end{itemize}

These systems highlight once more how important it is to unravel the complex interplay between group weather and AGN activity in the galaxy group regime. This is a fundamental step toward understanding the broader, multiphase mechanics of structure formation and evolution in the environments where most galaxies live. To fully assess the delicate balance between AGN feedback and radiative cooling, expanding these investigations to a long-term systematic study of nonthermal radio emission across a much larger sample will be essential. Future high-resolution, multifrequency radio observations (e.g., exploiting the full capabilities of the SKA or LOFAR2.0), combined with deeper X-ray spectroscopy from upcoming missions, will be crucial to accurately map magnetic fields, resolve low-Mach number shocks, and investigate the mechanisms that regulate the evolution of galaxy groups.

\begin{acknowledgements}
TP and FDG acknowledge support from the ERC Consolidator Grant ULU 101086378.
MG acknowledges support from the ERC Consolidator Grant \textit{BlackHoleWeather} (101086804). K Kolokythas acknowledges funding support from the South African Radio Astronomy Observatory (SARAO) and the National Research Foundation (NRF) (grant UID: 97930).
E. Bulbul acknowledge financial support from the European Research Council (ERC) Consolidator Grant under the European Union’s Horizon 2020 research and innovation program (grant agreement CoG DarkQuest No 101002585). The MeerKAT telescope is operated by the South African Radio Astronomy Observatory, which is a facility of the National Research Foundation, an agency of the Department of Science, Technology and Innovation. This research has made use of data obtained from the \textit{Chandra} Data Archive provided by the \textit{Chandra} X-ray Center (CXC). Based on observations obtained with \textit{XMM-Newton}, an ESA science mission with instruments and contributions directly funded by ESA Member States and NASA'. 

\end{acknowledgements}

\bibliographystyle{aa}
\bibliography{bibliography}

\begin{onecolumn}
\begin{appendix}

\section{Spectral index error maps}
\label{app:a}

We show below the spectral index error maps for NGC\,3411 and NGC\,5903, produced as described in Sect. \ref{sec:results}.

\begin{figure*}[!h]
\centering
  \includegraphics[scale=0.35]{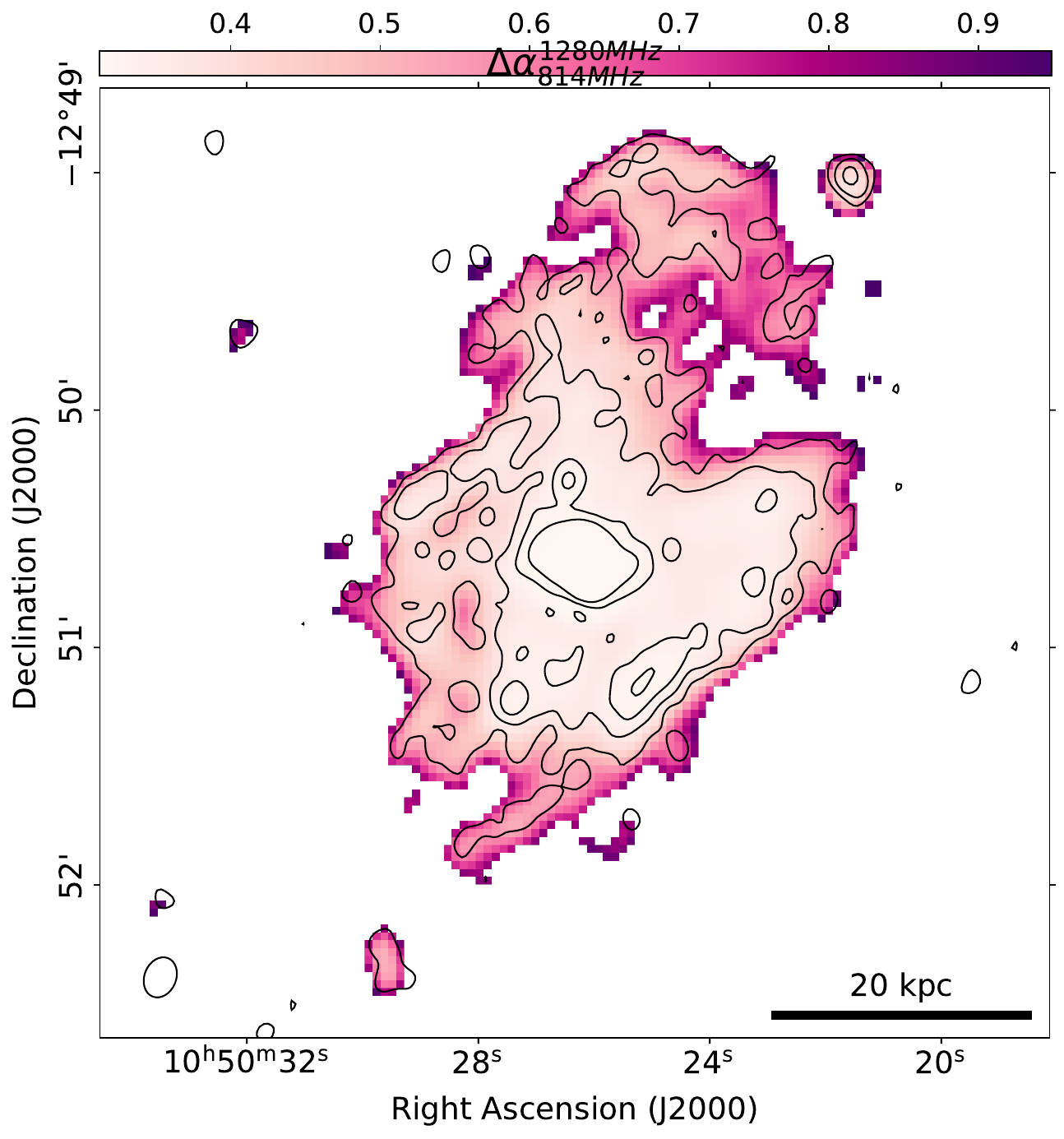}
  \hspace{1cm}
  \includegraphics[scale=0.35]{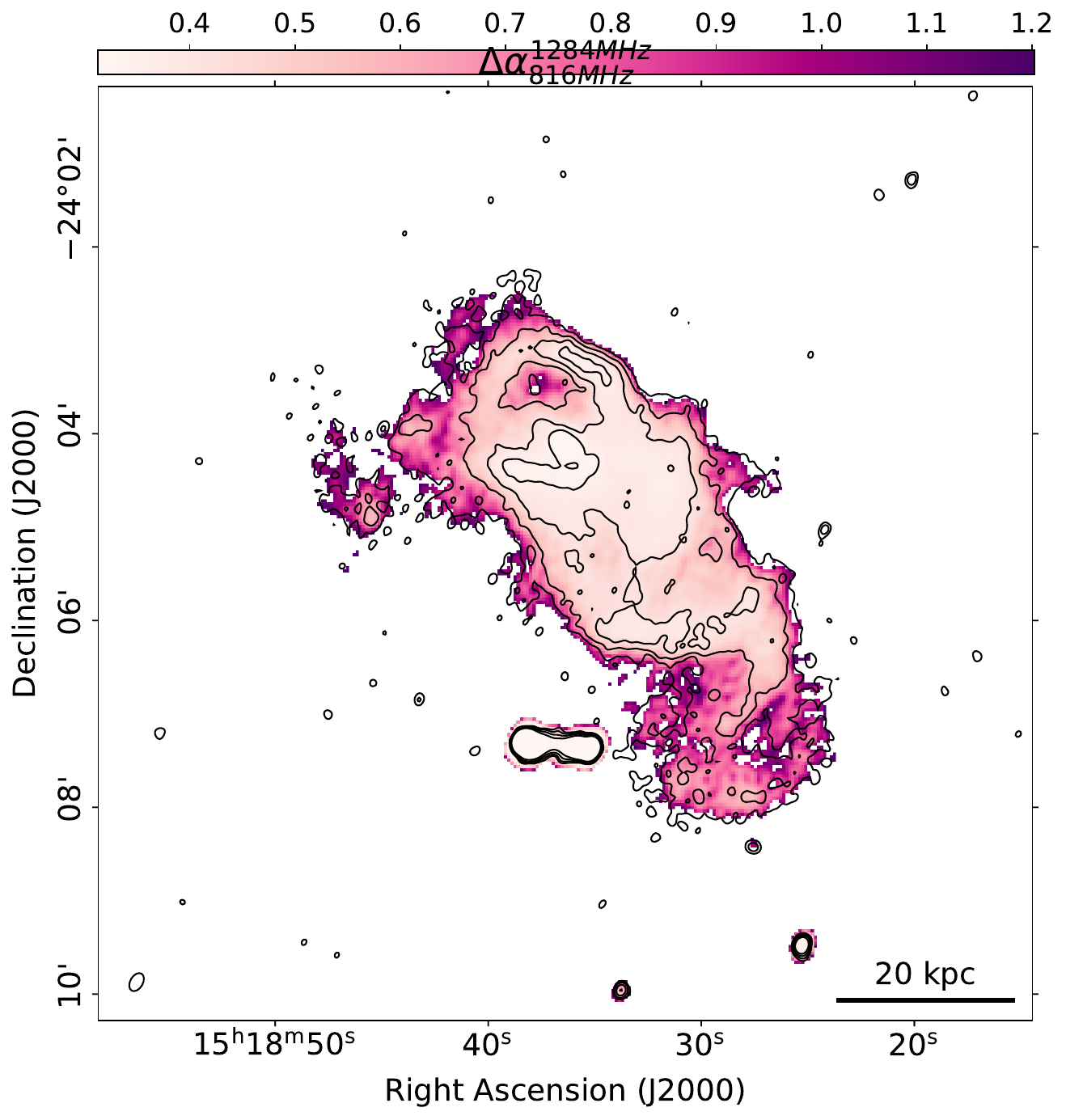}
  \caption{\textit{Left}: Spectral index error map of NGC\,3411. \textit{Right}: Spectral index error map of NGC\,5903.}
  \label{fig:spidxerr}
\end{figure*}

\section{Polarization maps of NGC\,5903}

\label{app:b}
\begin{figure*}[h!]
\centering
  \includegraphics[scale=0.35]{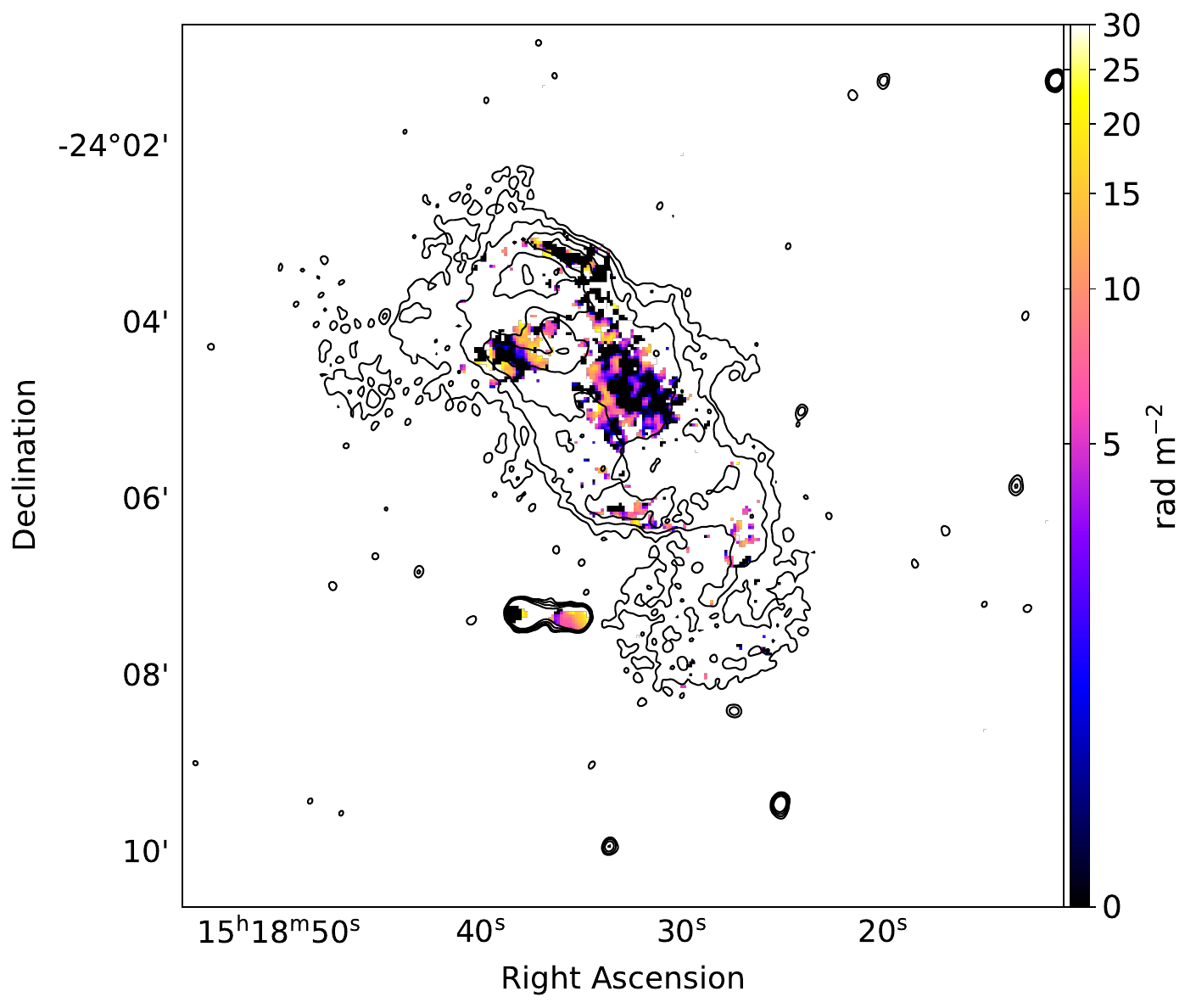}
  \hspace{0.3cm}
  \includegraphics[scale=0.35]{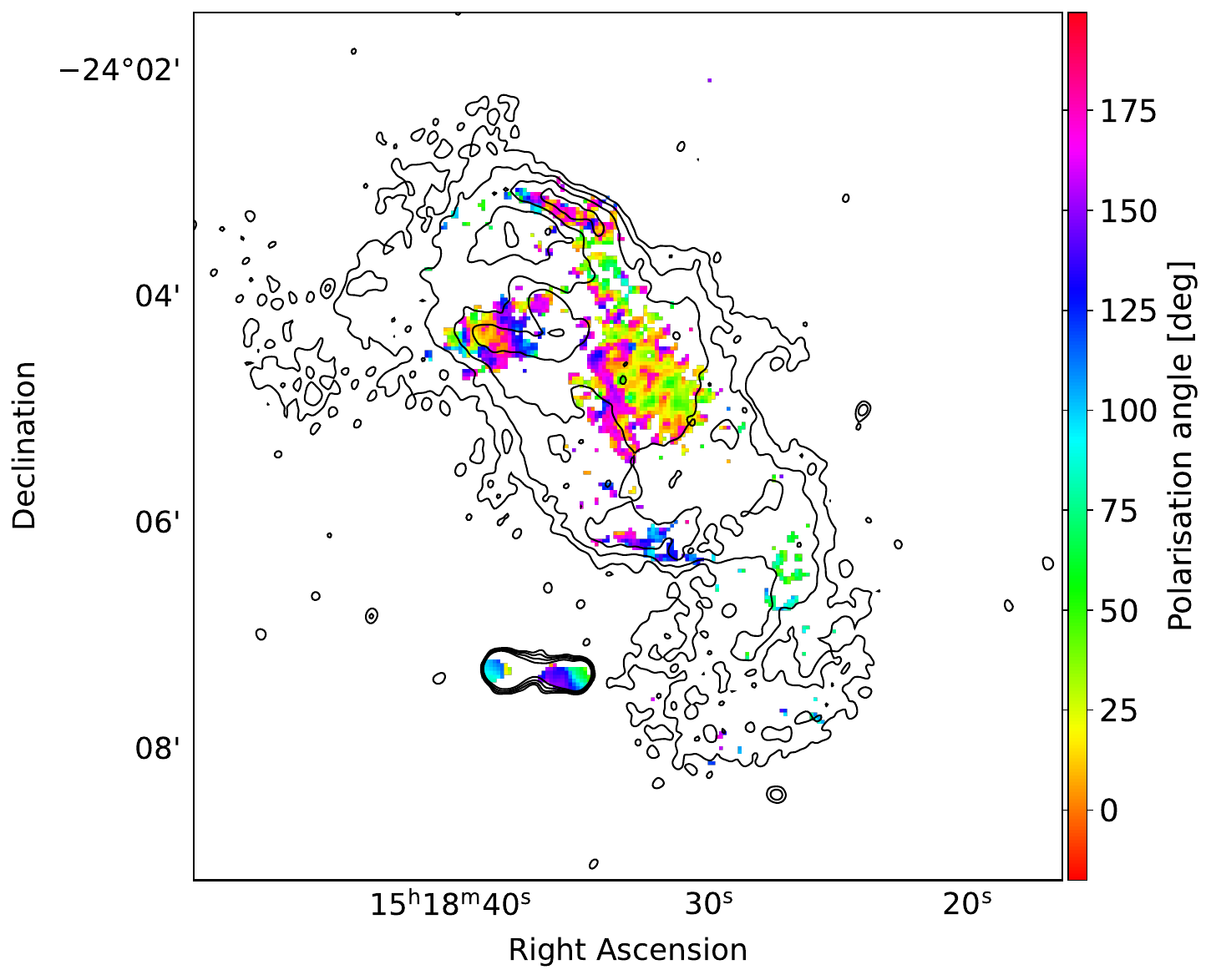}
  \caption{Polarization maps in the L band of NGC\,5903. \textit{Left}: RM map, clipped at 6$\sigma$, with total intensity contours overlaid in black. \textit{Right}: Polarization angle map, clipped at 6$\sigma$, with total intensity contours overlaid in black. }
  \label{fig:NGC5903_polarisation_rest}
\end{figure*}

We show above the RM and polarization angle map of NGC\,5903, produced as described in Sect. \ref{sec:results}. The patchy RM distribution observed in the left panel suggests an inhomogeneous magneto-ionic screen in the group plasma. RM values are close to zero, with localized peaks at $\sim$20, which argues against a high IGrM density as such low values imply a modest line-of-sight Faraday depth.\footnote{We note that contamination from Galactic foreground should not significantly affect this. For example, the S-band Polarisation All-Sky Survey (S-PASS; \citealt{Carretti_2019}) reports RM values of $\sim$-7 in this sky region, with typical uncertainties of $\sim$2.}

\end{appendix}

\end{onecolumn}

\end{document}